\RequirePackage{lineno}

\documentclass[aps, prl, preprint, nofootinbib, superscriptaddress, showpacs]{revtex4-1}
\usepackage{graphicx}
\usepackage{dcolumn}
\usepackage{overpic}
\usepackage{bm}
\usepackage{color}
\usepackage{mathrsfs}
\usepackage{amsmath}
\usepackage{amssymb}
\usepackage{graphicx}
\usepackage{url}
\usepackage[colorlinks=false]{hyperref}
\usepackage[textsize=footnotesize]{todonotes}
\usepackage[todos]{CMImacros}
\usepackage{pdfpages}

\begin{document}
\title{Tomographic imaging of complete quantum state of matter by ultrafast diffraction}
\author{Ming Zhang*}
\affiliation{State Key Laboratory for Mesoscopic Physics and Collaborative Innovation Center of Quantum Matter, School of Physics, Peking University, Beijing 10087, China}
\author{Shuqiao Zhang*}
\affiliation{State Key Laboratory for Mesoscopic Physics and Collaborative Innovation Center of Quantum Matter, School of Physics, Peking University, Beijing 10087, China}
\author{Haitan Xu*}
\affiliation{Institute for Quantum Science and Engineering and Department of Physics, Southern University of Science and Technology, Shenzhen, 518055, China}
\author{Hankai Zhang}
\affiliation{State Key Laboratory for Mesoscopic Physics and Collaborative Innovation Center of Quantum Matter, School of Physics, Peking University, Beijing 10087, China}
\author{Xiangxu Mu}
\affiliation{State Key Laboratory for Mesoscopic Physics and Collaborative Innovation Center of Quantum Matter, School of Physics, Peking University, Beijing 10087, China}
\author{R. J. Dwayne Miller}
\affiliation{Departments of Chemistry and Physics, University of Toronto, Toronto, Ontario M5S 3H6, Canada}
\author{Anatoly A. Ischenko}
\affiliation{Lomonosov Institute of Fine Chemical Technologies, RTU-MIREA - Russian Technological University, Vernadskii Avenue 86, 119571 Moscow, Russia}
\author{Oriol Vendrell}
\affiliation{Physikalisch-Chemisches Institut, Universit\"at Heidelberg, Im Neuenheimer Feld 229, D-69120 Heidelberg, Germany}
\author{Zheng Li}
\email{zheng.li@pku.edu.cn}
\affiliation{State Key Laboratory for Mesoscopic Physics and Collaborative Innovation Center of Quantum Matter, School of Physics, Peking University, Beijing 10087, China}

\date{\today}
\def\ket#1{|#1\rangle}
\def\Bra#1{\langle#1|}
\def\IP#1#2{\langle#1|#2\rangle}
\def\BK#1#2#3{\langle#1|#2|#3\rangle}
\def\ketBra#1#2{|#1\rangle\langle#2|}
\def\denmat#1#2{\langle#1|\hat{\rho}|#2\rangle}
\def\Pr{\ensuremath{\text{Pr}}}



\definecolor{oldtxtcolor}{rgb}{0.00, 0.0, 0.5}
\definecolor{newtxtcolor}{rgb}{0.00, 0.3867, 0.00}
\definecolor{newtxtcolor}{rgb}{0.00, 0.0, 1}
\definecolor{oldtxtcolor}{rgb}{1.00, 0.0, 0.00}

\def\verX{12}
\def\verO{1}
\def\verN{2}
\def\verON{12}

\ifx\verX\verO
 \newcommand { \oldtxt }[1] {{\color{oldtxtcolor}{#1}}}
 \newcommand { \newtxt }[1] {}
\fi
\ifx\verX\verN
 \newcommand { \oldtxt }[1] {}
 \newcommand { \newtxt }[1] {{\color{newtxtcolor}{#1}}}
\fi
\ifx\verX\verON
 \newcommand { \oldtxt }[1] {{\color{oldtxtcolor}{#1}}}
 \newcommand { \newtxt }[1] {{\color{newtxtcolor}{#1}}}
\fi

\maketitle

\begin{figure*}
    \centering
    \includegraphics[width=12.0cm]{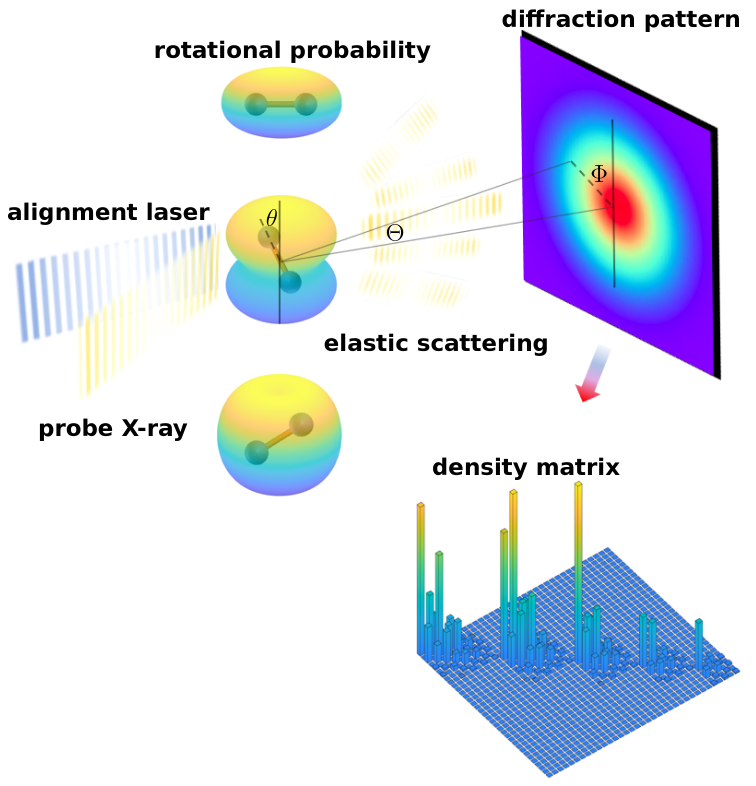}
    \caption{\label{fig:overview}
    Schematic drawing of quantum tomography by ultrafast diffraction, illustrated with a rotational wavepacket of N$_2$ molecule. A rotational wavepacket is prepared by an impulsive alignment laser pulse~\cite{Stapelfeldt03:RMP543}, and probed by diffraction of an incident X-ray pulses for a series of time intervals. The mixed rotational quantum state represented by its density operator $\hat{\rho}$ is determined from the diffraction patterns.}
\end{figure*}

\textbf{
With the ability to directly obtain the Wigner function and density matrix of photon states, quantum tomography (QT) has had a significant impact on quantum optics~\cite{Lvovsky09:RMP299, Priebe17:NatPho11, Smithey93:PRL1244}, quantum computing~\cite{Jianwei12:Science363,Laflamme20:Nat59} and quantum information~\cite{Murch13:Nat214,saglamyurek15:83}.
By an appropriate sequence of measurements on the evolution of each degree of freedom (DOF), the full quantum state of the observed photonic system can be determined.
The first proposal to extend the application of QT to reconstruction of complete quantum states of matter wavepackets~\cite{Leonhardt96:PRL1985} had generated enormous interest in ultrafast diffraction imaging~\cite{Yangjie18:Sci64, Wolf19:NatChem504, ischenko17:CR11066,gao13:343,eichberger10:799,sciaini09:56,mehrabi19:1167,ishikawa15:1501,miller14:6175,ernstorfer09:5917,siwick03:302,wolter16:308} and pump-probe spectroscopy of molecules~\cite{Dunn95:PRL884}.
This interest was elevated with the advent of ultrafast electron and X-ray diffraction techniques using electron accelerators and X-ray free electron lasers to add temporal resolution to the observed nuclear and electron distributions. In this respect, quantum tomography holds great promise to enable imaging of molecular wavefunctions beyond classical description.
This concept could become a natural area for quantum tomography of quantum states of matter~\cite{yang12:PRL133202,Cosmin12:Nat,YangJie16:Nat11232,Fielding18:Science30,Li19:AP296,YangJie16:Nat11232}.
However, the great interest in this area has been tempered by the illustration of an "impossibility theorem", known as the dimension problem~\cite{Mouritzen05:PRA, Mouritzen06:JCP244311}.
To obtain the density matrix of a system, the established QT procedure relies on integral transforms (e.g. the tomographic Radon transform), which preserves dimensionality~\cite{Lvovsky09:RMP299}.
Unlike its quantum optics sibling, only a single evolutionary parameter, time, is available for the molecular wavepacket.
Not being able to associate unitary evolution to every DOF of molecular motion, quantum tomography could not be used beyond 1D and categorically excludes most vibrational and all rotational motion of molecules.
Here we present a theoretical advance to overcome the notorious dimension problem.
Solving this challenging problem is important to push imaging molecular dynamics to the quantum limit.
The new theory has solved this problem, which makes quantum tomography a truly useful methodology in ultrafast physics and enables the making of quantum version of a ``molecular movie"~\cite{Nango16:Science1552, Fielding18:Science30, Li19:AP296, Nicholson18:Science821, Weinstein12:Nat157, gao13:343,miller14:6175,Yang20:Sci885}.
With the new theory, quantum tomography can be finally advanced to a sufficient level to become a general method for reconstructing quantum states of matter, without being limited in one dimension.
{Our new concept is demonstrated} using a simulated dataset of ultrafast diffraction experiment of laser-aligned nitrogen molecules~\cite{YangJie16:Nat11232}. The analysis with the new method reveals the density matrix of the rotational wavepacket (schematically shown in Fig.~\ref{fig:overview}), which is otherwise impossible to obtain with previously established QT procedures.
We also show that our approach can be naturally applied to quantum tomography of vibrational states to cover the complete DOF of molecular motion, and provides the ultimate information we can retrieve about dynamics of molecules from a quantum perspective. This approach can be potentially used in quantum computing and quantum information whenever quantum state information is tainted by insufficient evolutionary dimensions or incomplete measurements.
}
The modern formulation of quantum tomography based on integral transform~\cite{Lvovsky09:RMP299, Leonhardt96:PRL1985, Dunn95:PRL884} conceals its underlying nature as a retrieval procedure of wavefunction phases lost in the measurement.
Dating back to 1933, Pauli and Feenberg proposed that a wavefunction $\psi(x,t)=|\psi(x,t)|e^{i\phi(x,t)}$ can be obtained by measuring the evolution of 1D position probability distribution $\Pr(x,t)=|\psi(x,t)|^2$ and its time derivative $\partial \Pr(x,t)/\partial t$ for a series of times ~\cite{pauli33}. 
Equivalently, a pure quantum state can also be recovered by measuring $\Pr(x,t)$ at time $t$ and monitoring its evolution over short time intervals, i.e. $\Pr(x,t+N\Delta t)=|\psi(x,t+N\Delta t)|^2$ for $(N=0,1,2,\cdots)$.
{Reconstructing the phase of wavefunction} can be considered as the origin of quantum tomography. For a system with Hamiltonian $\hat{H}=\hat{H}_0+\hat{H}_{\text{int}}$, the established 1D QT method makes use of knowledge of the non-interacting part of the Hamiltonian $\hat{H}_0$, so that its eigenfunctions can be pre-calculated and used in the tomographic reconstruction of density matrix through integral inversion transform. However, for higher dimensional QT, the dimension problem as demonstrated in the pioneering works~\cite{Mouritzen05:PRA, Mouritzen06:JCP244311} mathematically leads to singularity in the inversion from the evolving probability distribution to the density matrix, {namely, failure of quantum tomography}.

{The fundamental principle of our approach is simple.} We solve the QT dimension problem by exploiting the interaction Hamiltonian $\hat{H}_{\text{int}}$ and the analogy between QT and crystallographic phase retrieval (CPR)~\cite{Rousse01:RMP17} in a seemingly distant field, crystallography.  
Further exploiting the interaction Hamiltonian $\hat{H}_{\text{I}}$ provides us a set of physical conditions, such as the selection rules of transitions subject to $\hat{H}_{\text{I}}$ and symmetry of the system. These physical conditions can be imposed as constraints in our QT approach, which is not feasible in the established QT methods based on integral transform.
By compensating with the additional physical conditions as constraints in the iterative QT procedure, the converged solution can be obtained as the admissible density matrix that complies with all the intrinsic properties of the investigated physical system.

We start by presenting the correspondence between QT and CPR. The research on CPR has been the focus of crystallography for decades~\cite{Rousse01:RMP17,Chapman11:Nat73,Seibert11:Nat78,Yangjie18:Sci64,Yang20:Sci885,yang12:PRL133202}. In crystallography, the scattered X-ray or electron wave encodes the structural information of molecules.
The measured X-ray diffraction intensity is
\begin{eqnarray}
I(\textbf{Q})\sim |F(\textbf{Q})|^2
\,,
\end{eqnarray}
where $\textbf{Q}=\textbf{k}_{f}-\textbf{k}_{\text{in}}$ is momentum transfer between incident and diffracted X-ray photon or electron, $F(\textbf{Q})$ is the electronically elastic molecular form factor.
For X-ray diffraction, the form factor is connected to the electron density by a Fourier transform 
$F_{\text{X}}(\textbf{Q}) \sim \mathscr{F}[\Pr(\textbf{x})]$,
$\Pr(\textbf{x})$ is the probability density of electrons in a molecule, and $\textbf{x}$ is the electron coordinate.
The form factor of electron diffraction has a similar expression $F_{\text{e}}(\textbf{Q})=[\Sigma_{\alpha} N_{\alpha}\exp(i\textbf{Q}\cdot\textbf{R}_{\alpha})-F_{\text{X}}(\textbf{Q})]/Q^2$, where $N_{\alpha}$, $\textbf{R}_{\alpha}$ are the charge and position of $\alpha^{\text{th}}$ nucleus.
However, the phase of the form factor, which is essential for reconstructing the molecular structure, is unknown in the diffraction experiment, only the modulus $|F(\textbf{Q})|$ can be obtained from measured diffraction intensity.

\begin{figure*}
    \centering
    \includegraphics[width=16.0cm]{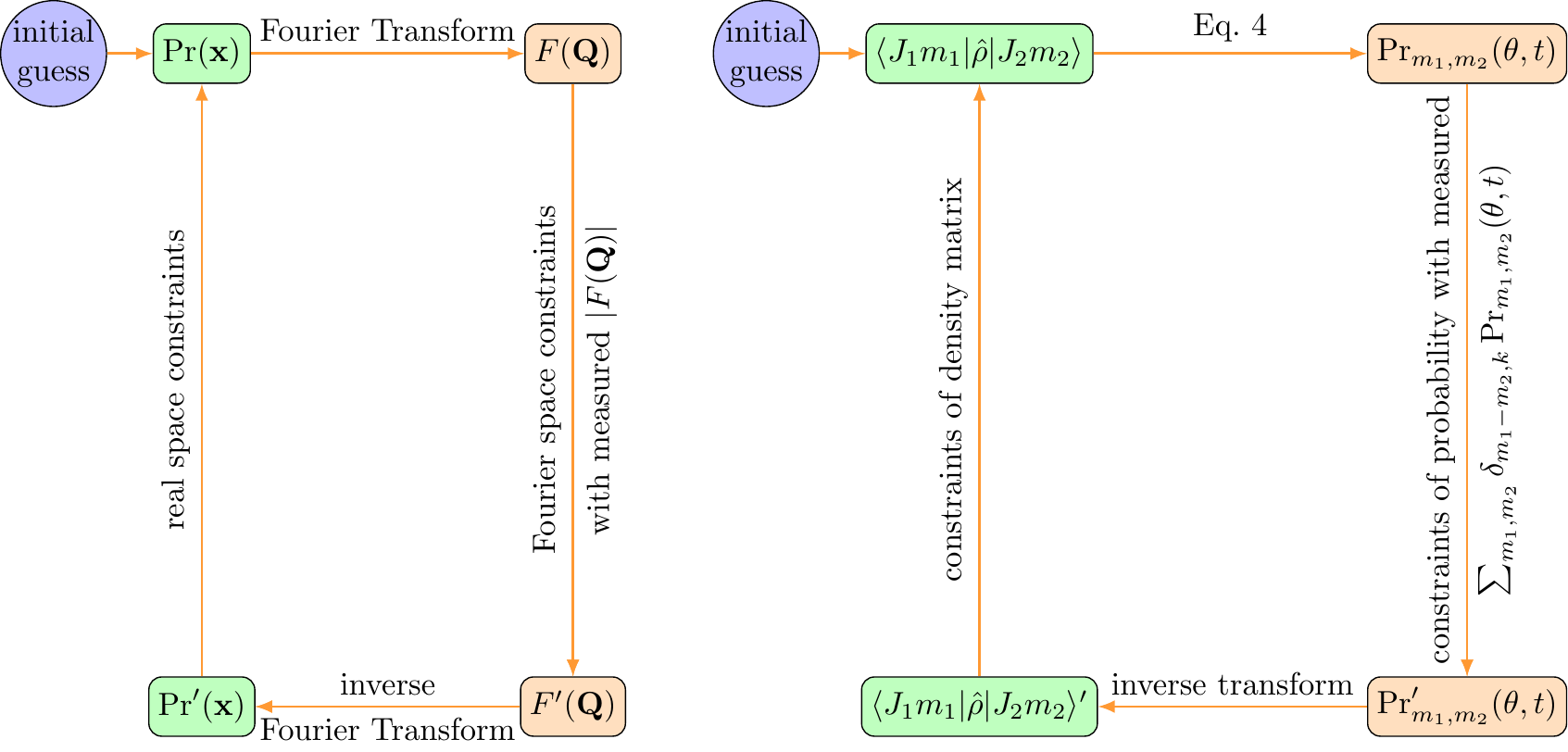}
    \caption{\label{fig1:analogy}
    Analogy between crystallographic phase retrieval (CPR) and quantum tomography (QT) based on their common nature~\cite{pauli33}. The CPR iterative transform between real space electron density $\Pr(\textbf{x})$ and Fourier space form factor $F(\textbf{Q})$ is analogously made for QT iterative transform between blockwise probability distribution $\Pr_{m_1,m_2}(\theta,t)$ in real space and elements in density matrix space.}
\end{figure*}

Phase retrieval is a powerful method that prevails in crystallography and single particle coherent diffraction imaging~\cite{yang12:PRL133202,Chapman11:Nat73,Seibert11:Nat78}.
Its basic idea is illustrated in Fig.~\ref{fig1:analogy}. Employing projective iterations between real space and Fourier space and imposing physical constraints in both spaces, the lost phases of the form factor $F(\textbf{Q})$ can be reconstructed with high fidelity. 
Fourier space constraint utilizes measured diffraction intensity data, and real space constraints comes from a priori knowledge, e.g. the positivity of electron density.
We present the new theory of quantum tomography based on this conceptual approach by applying it to a rotational wavepacket of nitrogen molecule prepared by impulsive laser alignment, using the ultrafast X-ray diffraction data.
Quantum tomography of rotational wavepackets is strictly impossible in the previously established QT theory, because the full quantum state of a rotating linear molecule is a 4D object $\BK{\theta,\phi}{\hat{\rho}}{\theta',\phi'}$, while the measured probability density evolution $\text{Pr}(\theta,\phi,t)$ is only 3D. It is obvious that the inversion problem to obtain the density matrix is not solvable by dimensionality-preserving transform.

From a dataset consisting of a series of time-ordered snapshots of diffraction patterns~\cite{HO08:PRA052409}
\begin{eqnarray}
I(\textbf{Q},t)=\int_0^{2\pi}d\phi\int_{0}^{\pi}\sin\theta d\theta \Pr(\theta,\phi,t) |F(\textbf{Q},\theta,\phi)|^2
\,,
\end{eqnarray}
the time-dependent molecular probability distribution $\text{Pr}(\theta,\phi,t)$ can be obtained by solving the Fredholm integral equation of the first kind (see Supplementary Information section 1 for details).
\begin{figure*}
    \centering
    \includegraphics[width=11.0cm]{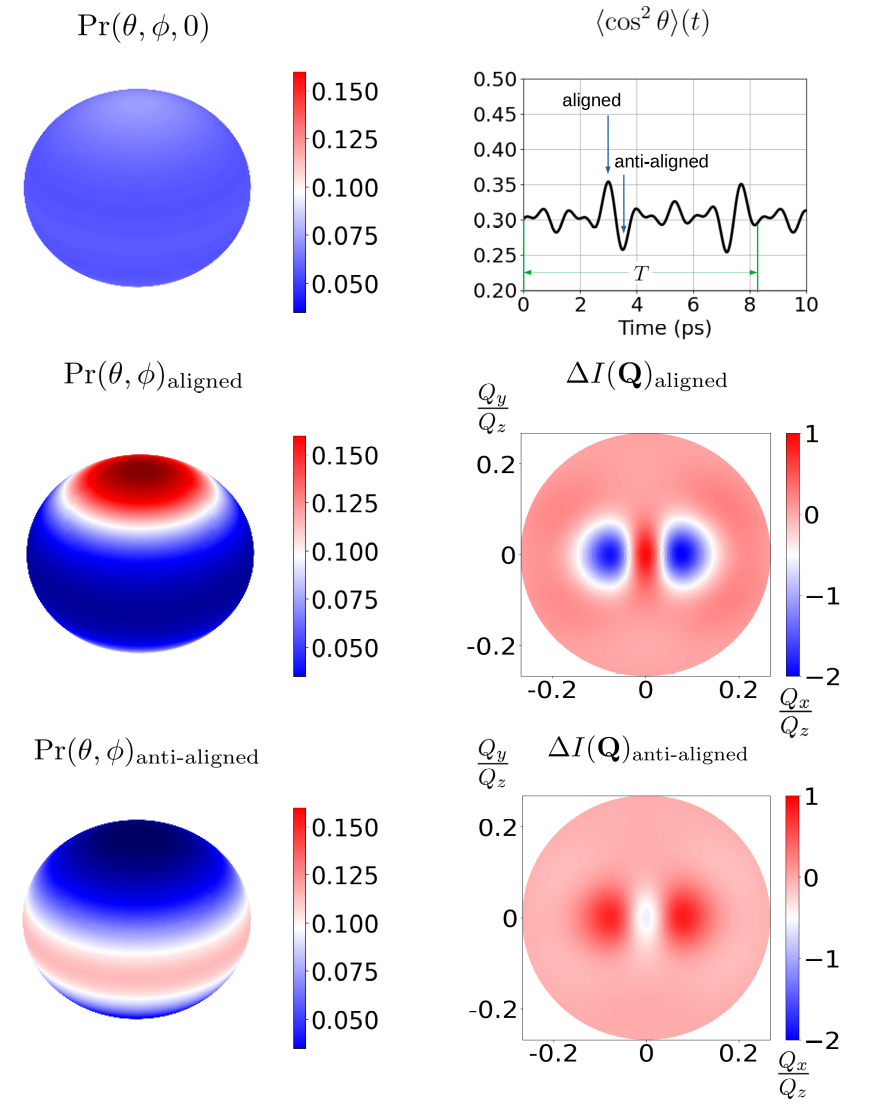}
    \caption{%
    Probability distribution and diffraction pattern of rotational wavepacket.
    The first row shows the initial angular probability for N$_2$ molecules prepared at rotational temperature of 30K and the expectation values of
     $\cos^2\theta$ of the time evolving wavepacket for N$_2$ molecules after laser pulse~\cite{Stapelfeldt03:RMP543}.
     The alignment laser pulse is linearly polarized with a Gaussian envelope of duration $\tau_L=50$fs and $10^{13}$W/cm$^2$ peak intensity, and $\theta$ is the polar angle between the polarization and the molecular axes. The duration is much shorter than the characteristic rotational time $\tau_L\ll T$.
     The second and third rows show the angular probability distribution changes from aligned to anti-aligned, and the difference of their diffraction intensity with respect to $t=0$ shown in Fig.~\ref{fig:overview}. The X-ray photon energy is assumed to be 20keV.}
\end{figure*}
%
The probability distribution of rotational wavepacket is
\begin{eqnarray}\label{Eq:den2pr}
    \Pr(\theta,\phi,t)
    = \sum_{J_1m_1}\sum_{J_2m_2}\BK{J_1m_1}{\hat{\rho}}{J_2m_2}   Y_{J_1m_1}(\theta,\phi)Y_{J_2m_2}^*(\theta,\phi)e^{-i\Delta\omega t}
    \,,
\end{eqnarray}
where $\Delta\omega=E_{J_1}-E_{J_2}$ is the energy spacing of rotational levels.
As shown in Fig.~\ref{fig1:analogy}, we devise an iterative procedure to connect the spaces of density matrix and temporal wavepacket density.

For the system of rotating molecules, the dimension problem limits the invertible mapping between density matrix and temporal wavepacket density to the reduced density of fixed projection quantum numbers $m_1$, $m_2$,
\begin{eqnarray}\label{Eq:mblock}
\Pr_{m_1,m_2}(\theta,t)=\sum_{J_1J_2}\denmat{J_1m_1}{J_2m_2}\tilde{P}_{J_1}^{m_1}(\cos\theta)\tilde{P}_{J_2}^{m_2}(\cos\theta)e^{-i\Delta\omega t}
\,.
\end{eqnarray}
The analytical solution of the inverse mapping from $\Pr_{m_1,m_2}(\theta,t)$ to density matrix $\denmat{J_1m_1}{J_2m_2}$ is elaborated in the Supplementary Information section 2.
However, due to the dimension problem, there is no direct way to obtain $\Pr_{m_1,m_2}(\theta,t)$ from the measured wavepacket density, only their sum is traceable through
$\sum_{m_1,m_2}\delta_{m_1-m_2,k} \Pr_{m_1,m_2}(\theta,t)
=\frac{1}{2\pi}\int_0^{2\pi} \Pr(\theta,\phi,t) e^{ik\phi} d\phi$.
%

%
Our method starts from an initial guess of density matrix and an iterative projection algorithm is used to impose constraints in the spaces of density matrix and spatial probability density.
The initial guess of quantum state, 
$\hat{\rho}_{\text{ini}} = \sum_{J_0m_0}\omega_{J_0} \ketBra{J_0m_0}{J_0m_0}$, 
is assumed to be an incoherent state in the thermal equilibrium of a given rotational temperature, which can be experimentally determined~\cite{YangJie16:Nat11232}.
$\omega_{J_0} =\frac{1}{Z} g_{J_0} e^{-\beta E_{J_0}}$ is the Boltzmann weight, and $g_{J_0}$ represents the statistical weight of nuclear spin, for the bosonic $^{14}$N$_2$ molecule, $g_{J_0}$ is 6 for even $J_0$ (spin singlet and quintet) and 3 for odd $J_0$ (spin triplet).


%
In the probability density space, constraint is imposed by uniformly scaling each reduced density $\Pr_{m_1,m_2}(\theta,t)$ with the measured total density $\Pr(\theta,\phi,t)$. Constraints in the density matrix space enable us to add all known properties of a physical state to the QT procedure, which supply additional information to compensate the missing evolutionary dimensions.
The constraints contain general knowledge of the density matrix, i.e. the density matrix is positive semidefinite, Hermitian and with a unity trace.
Besides, the selection rules of the alignment laser-molecule interaction imply further constraints on physically nonzero $m$-blocks of the density matrix and invariant partial traces of density matrix elements subject to rotational quantum number $J$ (see Supplementary Information section 3 for details of the algorithm).



\begin{figure*}
    \centering
    \includegraphics[width=16.0cm]{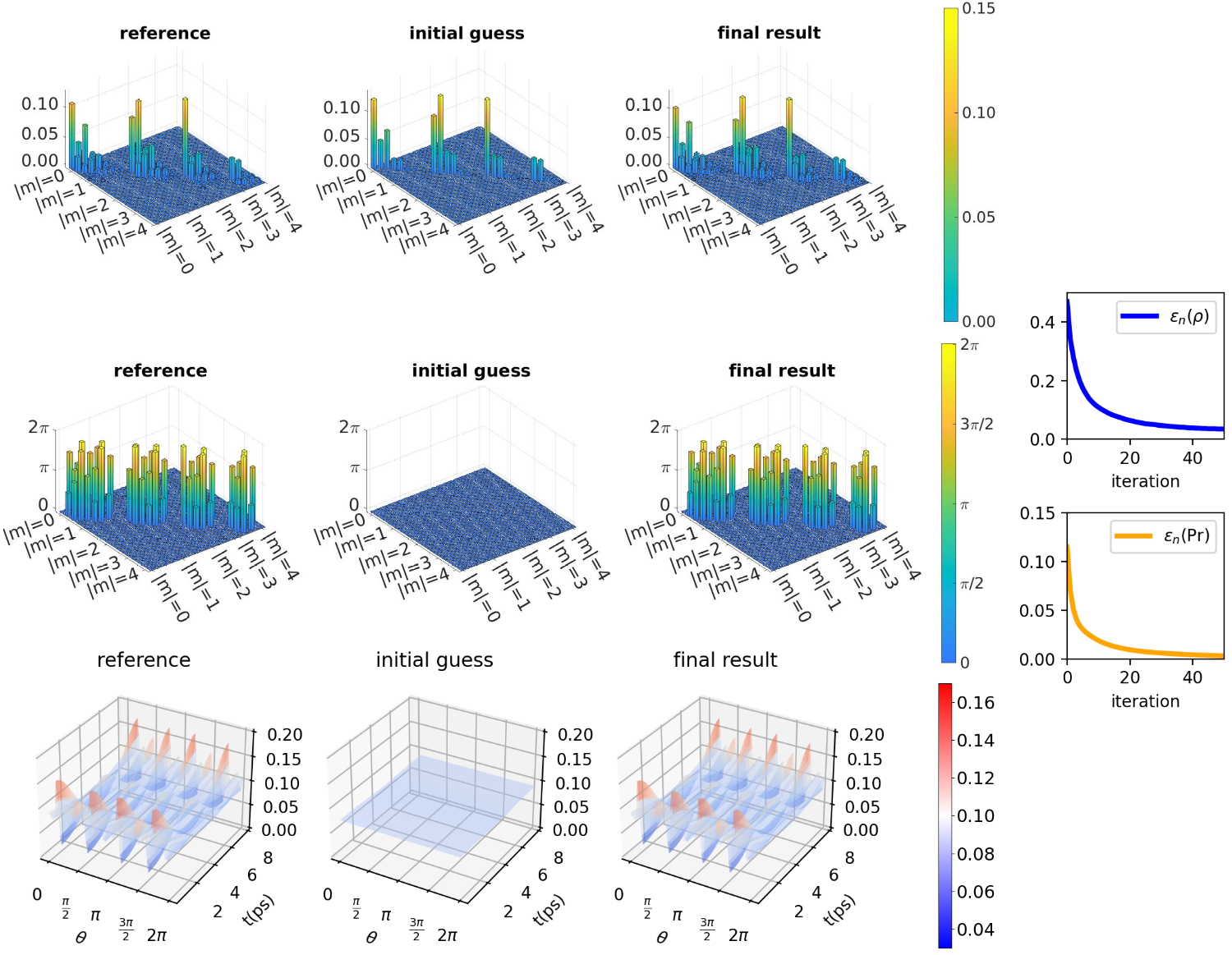}
    \caption{\label{fig3:qt}%
    Quantum tomography of rotational wavepacket of nitrogen molecule. 
    The modulus and phases of density matrix elements are shown in the upper and middle panel, within each $m$-block $J=|m|,|m|+1,\cdots,J_{\text{max}}$ (phases are at $t=0$). The density matrix element of opposite magnetic quantum number $m$ and $-m$ are identical (see Supplementary Information section 3).
    The lower panel shows the wavepacket probability distribution $\Pr(\theta,t)$.
    The convergence of the procedure is illustrated in the rightmost column.}
\end{figure*}
We demonstrate the capability of the new approach to recover the density matrix despite the dimension problem.
We use simulated ultrafast diffraction dataset of impulsively aligned nitrogen molecule, prepared at rotational temperature of 30 K.
The order of the recovered density matrix sets the requirement on the resolution. 
From Eq.~\ref{Eq:mblock}, the characteristic time scale of rotation is $\frac{2\mathcal{I}}{|\beta|(\alpha+1)}$, where $\mathcal{I}$ is the moment of inertia of nitrogen molecule, $\alpha$ and $\beta$ are two integral parameters satisfying $\beta(\alpha+1)=\Delta J(J+1)$ and $|\Delta J|\le |\beta| \le\alpha\le J$ where $\Delta J=J_1-J_2$ and $J=J_1+J_2$ for any two eigenstates with $J_1, J_2$ (see Supplementary Information section 2).
Using the Nyquist–Shannon sampling theorem, the required temporal resolution $\delta t$ should be  
$\delta t\leq\frac{\mathcal{I}}{|\beta|(\alpha+1)}$.
The spatial resolution $\delta \theta$ and $\delta \phi$ can be determined with the argument that the nodal structure of spherical harmonic basis in Eq.~\ref{Eq:den2pr} must be resolved, i.e. 
$\delta \theta<\frac{\pi}{2J_{\text{max}}}$.
%
%
To recover density matrix up to the order $J_{\text{max}}=5$, it demands time resolution $\delta t \lesssim 10^{2}$ fs and spatial resolution $\delta \theta
 \lesssim 10^{-1}$ rad.
Quantum tomography of the rotational wavepacket gives the result shown in Fig.~\ref{fig3:qt}. After 50 iterations, both density matrix and probability distribution are precisely recovered. 
The error of density matrix is $\epsilon_{50}(\hat{\rho})=2.9\times 10^{-2}$ and error of probability achieves $\epsilon_{50}(\Pr)=3.8\times 10^{-5}$.




\begin{figure*}
    \centering
    \includegraphics[width=12.0cm]{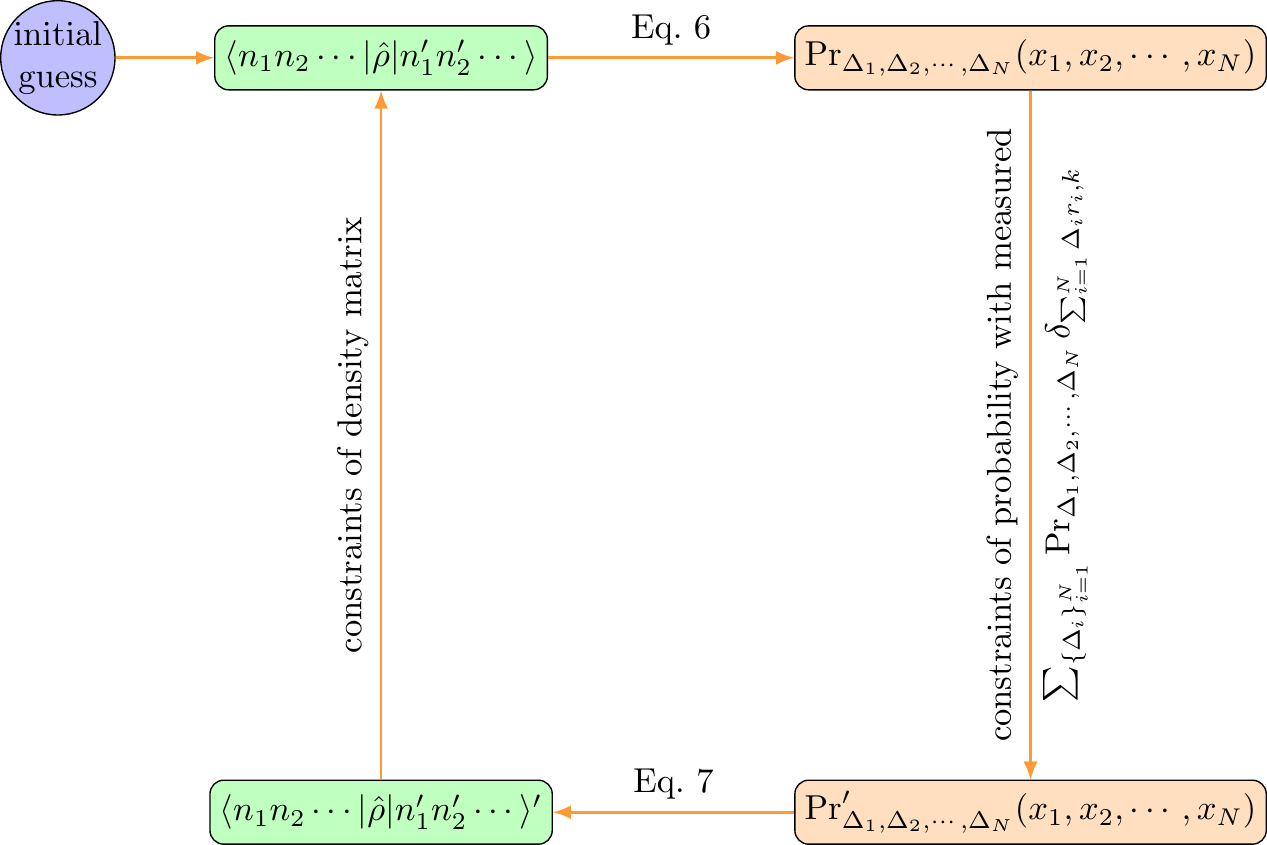}
    \caption{\label{fig4:vib}%
    Quantum tomography of vibrational state. The iterative transform is again between the spaces of density matrix and the blockwise probability distribution $\Pr_{\Delta_1,\Delta_2,\cdots,\Delta_N}(x_1,x_2,\cdots,x_N)$.}
\end{figure*}

The new quantum tomography procedure can be straightforwardly extended to obtain quantum state of vibrational wavepacket, which suffers from the dimension problem as well, when more than one vibrational modes are involved.
By ultrafast diffraction technique, we can measure the joint spatial probability density of $N$ vibrational modes for different time $t$~\cite{Ischenko14:Adv}. This gives an $(N+1)$-dimensional probability density $\Pr(x_1,x_2,\cdots,x_N,t)$, where $x_i$'s are the spatial coordinates of vibrational eigenmodes. The density matrix of the quantum state is however a $2N$-dimensional object $\denmat{n_1n_2\cdots n_N}{m_1m_2\cdots m_N}$ in the eigenmode basis, which cannot be obtained by dimensionality-preserving transform for $N>1$ (detailed proof is presented in Supplementary Information section 6).

The new approach provides a similar iterative quantum tomography procedure for vibrational wavepacket, with the capability to extract the vibrational density matrix $\denmat{n_1n_2\cdots n_N}{m_1m_2\cdots m_N}$ from the lower dimensional wavepacket density evolution $\Pr(x_1,x_2,\cdots,x_N,t)$.
The density matrix of vibrational wavepacket is to be recovered from the probability distribution
\begin{eqnarray}\label{denmat2prb}
    \Pr(x_1,x_2,\cdots,x_N,t)=\sum_{\{m_i\}_{i=1}^{N}}\sum_{\{n_i\}_{i=1}^{N}}\denmat{n_1n_2\cdots n_N}{m_1m_2\cdots m_N} \\\nonumber
    \times \prod_{i=1}^{N}\phi_{n_i}(x_i)\phi^*_{m_i}(x_i)e^{i(m_i-n_i)\omega_i t}
\end{eqnarray}
where $\omega_i$ and $\phi_{n_i}(x_i)$ are frequency and eigenstate of $n_i$-th vibrational mode.
Again, the dimension problem restricts the integral transformation between probability and density matrix in a subspace, in which the difference between two quantum numbers $\Delta_i$ of each vibrational mode is fixed
\begin{eqnarray}
    \Pr_{\Delta_1,\Delta_2,\cdots,\Delta_N}(x_1,x_2,\cdots,x_N)&=&\sum_{\{m_i\}_{i=1}^{N}}\sum_{\{n_i\}_{i=1}^{N}}\denmat{n_1n_2\cdots n_N}{m_1m_2\cdots m_N}\\\nonumber
    &&\times\prod_{i=1}^{N}\phi_{n_i}(x_i)\phi^*_{m_i}(x_i)\delta_{m_i-n_i,\Delta_i}\\
    \denmat{n_1n_2\cdots n_N}{m_1m_2\cdots m_N}&=&
    \int d^N\textbf{x} \Pr_{\Delta_1,\Delta_2,\cdots,\Delta_N}(x_1,x_2,\cdots,x_N)
    \times \prod_{i=1}^{N}f_{m_in_i}(x_i)
\end{eqnarray}
where the sampling functions $f_{mn}(x)$ are derivation of product of regular and irregular wavefunctions~\cite{Leonhardt96:PRL1985}, which is bi-orthogonal to both $\phi_{m}(x)$ and $\phi_{n}(x)$ provided the frequency constraint.
Then effective constraints can be imposed by iterative projection method to get the converged result satisfying physical condition in the density matrix and probability density spaces. For example, the information of measurable probability distribution makes up of the following constraint
\begin{eqnarray}
    \sum_{\{\Delta_i\}_{i=1}^{N}} \Pr_{\Delta_1,\Delta_2,\cdots,\Delta_N}(x_1,x_2,\cdots,x_N)\delta_{\sum_{i=1}^{N}\Delta_ir_i,k}=\frac{1}{T}\int_{0}^{T}dt e^{-ik\omega_0 t} \Pr(x_1,x_2,\cdots,x_N,t)
\end{eqnarray}
where we assume $\omega_i=r_i\omega_0$ ($r_i$ are integers and $T={2\pi}/{\omega_0}$).
To determine density matrix for vibrational wavepackets up to $N$th order, it demands measurement with $\delta{x}\leq \pi/2\sqrt{2N+1}$ and $\delta{t}\leq T/2(N+1)\Sigma_i r_i$(see Supplementary Information section 6). This leads to the requirement of spatial temporal resolution of $\delta{x}\sim 1$pm and $\delta{t}\sim 10$fs, which sets an ultimate demand for resolution in ultrafast diffraction.

The resemblance between the vibration and rotation problem proves the generality of our method, and sets up solid foundation to solve other quantum tomography problems as well as its application in relevant experiments.

In summary, we have demonstrated a new iterative quantum tomography approach that is capable of extracting the density matrix of high-dimensional wavepacket of matter from its evolutionary probability distribution in time.
The notorious dimension problem, which has prohibited for almost two decades the quantum tomographic reconstruction of molecular quantum state from ultrafast diffraction, has been finally resolved.
We expect this advance to have a broad impact in many areas of science and technology, not only for making the quantum version of molecular movie, but also for quantum state measurement in condensed matter physics, quantum computing and quantum information with imperfect knowledge of phase information, e.g. due to incomplete measurement for the sake of speed and efficiency.

\subsection{Acknowledgements}
We thank Jie Yang and Stefan Pabst for useful discussions. This work was supported by NSFC Grant No. 11974031, funding from state key laboratory of mesoscopic physics and RFBR Grant No. 20-02-00146.

\subsection{Author Contribution}
Z.L. designed the study. M.Z., S.Q.Z., H.K.Z. and Z.L. carried out the calculations. M.Z., S.Q.Z., H.K.Z., X.X.M., H.T.X., O.V., R.J.D.M., A.I. and Z.L. analysed the data. All authors contributed to the writing of the manuscript.

\bibliographystyle{naturemag}

\pagebreak


\section*{Supplementary material}

# SUPPLEMENTARY INFORMATION
# Tomographic imaging of complete quantum state of matter by ultrafast diffraction

## 1. ROTATIONAL PROBABILITY DISTRIBUTION FROM SOLVING THE FREDHOLM INTEGRAL EQUATION OF THE FIRST KIND

Facing the problem with one variable of integration, the integral equation is written as

$$I(\Theta) = \int_0^\pi |F(\Theta, \theta)|^2 \Pr(\theta) d\theta.$$

The integral equation can be approximately replaced by a Riemann summation over grids,

$$I(\Theta_j) = \sum_{k=1}^n |F(\Theta_j, \theta_k)|^2 \Pr(\theta_k) \Delta.$$

where $\Delta = \theta_{j+1} - \theta_j$. We can easily write the summation in the matrix form:

$$I = \mathbf{K} \Pr$$

where

$$I = \begin{pmatrix} I(\Theta_1) \\ \vdots \\ I(\Theta_j) \\ \vdots \\ I(\Theta_m) \end{pmatrix}, \mathbf{K} = \begin{pmatrix} |F(\Theta_1, \theta_1)|^2 \Delta & \cdots & |F(\Theta_1, \theta_n)|^2 \Delta \\ \vdots & \ddots & \vdots \\ |F(\Theta_m, \theta_1)|^2 \Delta & \cdots & |F(\Theta_m, \theta_n)|^2 \Delta \end{pmatrix}, \Pr = \begin{pmatrix} \Pr(\theta_1) \\ \vdots \\ \Pr(\theta_i) \\ \vdots \\ \Pr(\theta_n) \end{pmatrix} \quad (S1)$$

Assume $(\mathbf{K}^T \mathbf{K})^{-1}$ exists, and then the solution will be:

$$\Pr = (\mathbf{K}^T \mathbf{K})^{-1} \mathbf{K}^T I.$$

In many cases, the problems are ill-conditioned, and $\mathbf{K}^T \mathbf{K}$ is a singular matrix which is not invertible. So we have to use the technique of Tikhonov regularization [1].

Tikhonov regularization procedure aims to minimize $||\mathbf{K}\Pr - I||_2^2 + \lambda ||\Pr||_2^2$, where $\lambda$ is a control parameter. The second part is a penalty term in case of a divergent solution. And the solution of the minimization problem gives $\Pr = (\mathbf{K}^T \mathbf{K} + \lambda \mathbf{E})^{-1} \mathbf{K}^T I$, where $\mathbf{E}$ is identity matrix of size $m$.

When it comes to the integral equation with multiple variables, the same logic can be used. For rotationally averaged diffraction pattern of a linear molecule, the integral equation for diffraction intensity $I(\Theta, \Phi)$ from a given rotational probability distribution $\Pr(\phi, \theta)$ becomes [2, 3]

$$I(\Theta, \Phi) = \int_0^{2\pi} d\phi \int_0^{\pi} \sin\theta d\theta |F(\phi, \theta, \Theta, \Phi)|^2 \Pr(\phi, \theta), \qquad (S2)$$

where $\phi$ and $\theta$ are the azimuthal and levitation angles of the linear molecular rotor, $\Theta$ and $\Phi$ are the scattering angle of the X-ray photon in the lab system (as is shown in Fig. 1 in the main text), which gives the momentum transfer

$$\mathbf{Q} = 2(\omega_{\mathrm{in}}/c)\sin(\Theta/2) \begin{pmatrix} -\cos\Phi\cos(\Theta/2) \\ -\sin\Phi\cos(\Theta/2) \\ \sin(\Theta/2) \end{pmatrix}, \qquad (S3)$$

for incident photon frequency $\omega_{\mathrm{in}}$. $F(\phi, \theta, \Theta, \Phi)$ is the molecular form factor, which can be calculated for given molecular structure by the independent atom model (IAM) or ab initio electron density. In this work, IAM method is used to calculate the molecular form factor [2]. We assume $\tau = -\cos\theta$ and rewrite the integral equation as

$$I(\Theta, \Phi) = \int_0^{2\pi} d\phi \int_{-1}^{1} d\tau |F(\phi, \theta(\tau), \Theta, \Phi)|^2 \Pr(\phi, \theta(\tau)),$$

and replace the integral by Riemann summation,

$$I(\Theta_k, \Phi_l) = \sum_{i=1}^{a} \Delta\phi \sum_{j=1}^{b} \Delta\tau |F(\phi_i, \theta(\tau_j), \Theta_k, \Phi_l)|^2 \Pr(\phi_i, \theta(\tau_j)),$$

where $\Delta\phi = \frac{2\pi}{a}$, $\Delta\tau = \frac{2}{b}$, $i$ is ranging from 1 to $a$, $j$ is ranging from 1 to $b$, $k$ is ranging from 1 to $c$, and $l$ is ranging from 1 to $d$. We can write the total diffraction intensity in the matrix form:

$$I = \mathbf{K}\Pr,$$

where

$$I = \begin{pmatrix} I(\Theta_1, \Phi_1) \\ I(\Theta_1, \Phi_2) \\ \vdots \\ I(\Theta_1, \Phi_d) \\ I(\Theta_2, \Phi_1) \\ \vdots \\ I(\Theta_k, \Phi_l) \\ \vdots \\ I(\Theta_c, \Phi_d) \end{pmatrix}, \mathbf{K} = \begin{pmatrix} |F(\phi_1, \theta_1, \Theta_1, \Phi_1)|^2 \Delta\phi\Delta\tau & \cdots & |F(\phi_a, \theta_b, \Theta_1, \Phi_1)|^2 \Delta\phi\Delta\tau \\ \vdots & \ddots & \vdots \\ |F(\phi_1, \theta_1, \Theta_c, \Phi_d)|^2 \Delta\phi\Delta\tau & \cdots & |F(\alpha_a, \theta_b, \Theta_c, \Phi_d)|^2 \Delta\phi\Delta\tau \end{pmatrix}, \mathrm{Pr} = \begin{pmatrix} \rho(\phi_1, \theta_1) \\ \rho(\phi_1, \theta_2) \\ \vdots \\ \rho(\phi_1, \theta_b) \\ \rho(\phi_2, \theta_1) \\ \vdots \\ \rho(\phi_a, \theta_b) \end{pmatrix}$$

(S4)

To avoid singular matrix inversion, we use Tikhonov regularization to get the rotational probability distribution,

$$\mathrm{Pr} = (\mathbf{K}^T \mathbf{K} + \lambda \mathbf{E})^{-1} \mathbf{K}^T I, \tag{S5}$$

where $\mathbf{E}$ is identity matrix of size $(c \times d)$ and $\mathbf{K}^T$ is the transpose of matrix $\mathbf{K}$.

It is important to validate the faithfulness of the obtained probability distribution $\mathrm{Pr}(\theta, \phi)$, i.e. the solution must be sufficiently close to the reality despite of regularization of the possibly singular matrix. For this purpose, we show here that the probability distribution is insensitive to the measurement error of diffraction intensity as well as the choice of regularization parameter. Especially, in practice we can use the condition number of the integral equation as the a priori criterion to check the reliability of the solution, which is defined as

$$\mathrm{cond} = \frac{\|\Delta\mathrm{Pr}\|_2 / \|\mathrm{Pr}\|_2}{\|\Delta I\|_2 / \|I\|_2},$$

where $\|A\|_2 = \sqrt{\sum_i A_i^2}$ is the $\mathcal{L}^2$ Euclid norm. The condition number characterizes the degree of variation of the solution $\mathrm{Pr}(\theta, \phi)$ with respect to the input data of measured diffraction intensity $I(\mathbf{Q})$, its value provides a measure for the sensitivity of the solution with respect to the measurement error and choice of regularization parameters. A problem with a small condition number is considered to be well-conditioned.

For our specific problem, we generate $I' = I + \Delta I$ by adding a small random part to all elements of $I$, where $\Delta I_i = 0.01 \cdot \mathrm{rand}A \cdot |I_i|$ and $\mathrm{rand}A$ is a random number ranging

from $-1$ to $1$. We then use the technique of Tikhonov regularization to get $\mathrm{Pr}'$ and define $\Delta\mathrm{Pr} = \mathrm{Pr}' - \mathrm{Pr}$. With $\Delta\mathrm{Pr}$, $\mathrm{Pr}$, $\Delta I$, $I$ obtained, we can calculate the corresponding condition number.

For a well-conditioned problem with small condition number, even a big change in $I$ will not lead to great change in Pr, namely, the wavepacket probability distribution Pr is insensitive to the measurement error of diffraction intensity $I$. We choose here cond $\lesssim 10$ as the criterion for well-conditioned solutions. From Fig. S1, we can estimate that $\lambda \gtrsim 10$ is required to ensure cond $\lesssim 10$, and subsequently to ensure the reliability of the solution. Also, the regularization parameter $\lambda$ cannot be chosen too large, otherwise the problem we solve will have little connection with the original problem. From right panel of Fig. S1 we can see as $\lambda$ grows larger, $\|\mathrm{Pr}\|_2$ decreases and the residual $\|I - \mathbf{K} \cdot \mathrm{Pr}\|_2$ increases. For a stable solution, Pr should be at an interval so that $\|\mathrm{Pr}\|_2^2$ does not change a lot as $\lambda$ varies, so we demand the ratio of the residual $\|I - \mathbf{K} \cdot \mathrm{Pr}\|_2^2$ and $\|I\|_2^2$ to be less than $10^{-5}$. From Fig. S1 we find that $\lambda < 10^4$ is required. From the above procedure, we have determined a priori the appropriate interval of regularization parameter to be $\lambda \in [10, 10^4]$, which ensures both the faithfulness and the accuracy of the obtained probability distribution.

Our problem is solved with $\lambda = 10^2$ by the technique of Tikhonov regularization for diffraction intensities throughout the period. All equations have condition number less than 10, which proves that they are stable and well-conditioned, because for cond $\leq 10^1$, we can expect to lose less than only 1 digit of precision in numerically solving the integral equation. The relative residual $\frac{\|I - K \cdot \mathrm{Pr}\|_2^2}{\|I\|_2^2}$ of all solutions are less than $10^{-7}$. Since both the residual $\Delta I = I - \mathbf{K} \cdot \mathrm{Pr}$ and the condition number are small, we can infer that $\Delta\mathrm{Pr}$ is very small and our result is faithful to the real value of the rotational probability distribution Pr.

## 2. QUANTUM TOMOGRAPHY FOR STATES IN $m$-BLOCK WITH FIXED PROJECTION QUANTUM NUMBER

We extend the treatment in Ref. [4] to show that the density matrix element $\langle J_1 m_1 | \hat{\rho} | J_2 m_2 \rangle$ in the $(m_1, m_2)$-block subspace can be solved analytically, once the blockwise probability density $\mathrm{Pr}_{m_1, m_2}(\theta, t)$ of given projection quantum number $m_1, m_2$ is determined. We expand

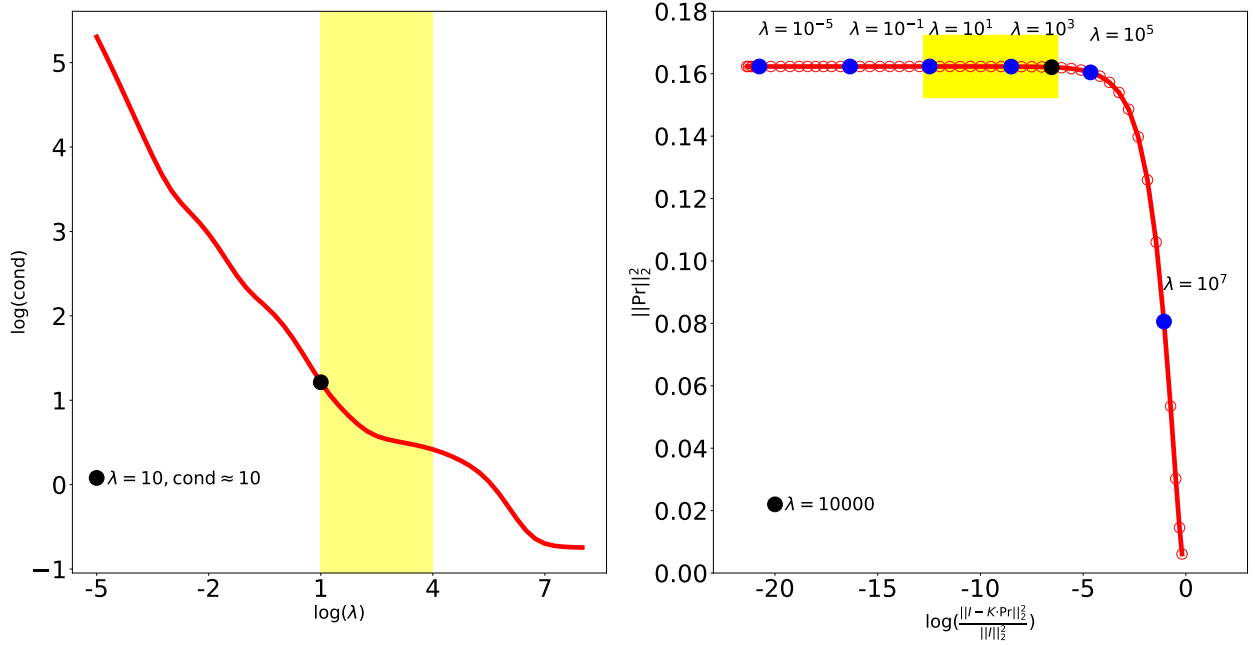

FIG. S1. Faithfulness of the probability distribution Pr obtained from integral equation with Tikhonov regularization. (Left) Logarithm of condition number versus logarithm of the regularization parameter $\lambda$. Larger $\lambda$ makes the problem more insensitive to the measurement error $\Delta I$. The approximate position of the black point marked on the sketch is (1,1) (we use an approximate position because every calculation that contains generation of the random numbers leads to slightly different curve). (Right) The values of $\|\mathrm{Pr}\|_2^2$ and the residual $\log(\frac{\|I-\mathbf{K}\cdot\mathrm{Pr}\|_2^2}{\|I\|_2^2})$ for $\lambda$ ranging from $10^{-5}$ to $10^8$. The Tikhonov regularization procedure minimizes $\|I - \mathbf{K}\mathrm{Pr}\|_2^2 + \lambda\|\mathrm{Pr}\|_2^2$. The black point marked on the curve is the turning point corresponding to $\lambda \approx 10^4$. The yellow area starting from $\log\lambda = 1$ and ending at $\log\lambda = 4$ illustrates the admissible range of regularization parameter $\lambda$.

the blockwise probability density with eigenbasis,

$$\mathrm{Pr}_{m_1,m_2}(\theta, t) = \sum_{J_1=|m_1|}^{\infty} \sum_{J_2=|m_2|}^{\infty} \langle J_1 m_1|\hat{\rho}|J_2 m_2\rangle \widetilde{P}_{J_1}^{m_1}(\cos\theta)\widetilde{P}_{J_2}^{m_2}(\cos\theta)e^{-i\Delta\omega t}, \tag{S6}$$

where the energy level difference is

$$\Delta\omega = E_{J_1} - E_{J_2} = \frac{\Delta J(J+1)}{2\mathcal{I}},$$

5

$\Delta J = J_1 - J_2$, $J = J_1 + J_2$ and $\mathcal{I}$ is the moment of inertia of the rotating molecule. For the sake of convenience, we define normalized associated Legendre polynomials

$$\widetilde{P}_j^m(\cos\theta) = (-1)^m \sqrt{\frac{(2j+1)(j-m)!}{2(j+m)!}} P_j^m(\cos\theta), \quad (S7)$$

with orthonormal relations

$$\int_0^\pi \sin\theta d\theta \widetilde{P}_{j_1}^m(\cos\theta)\widetilde{P}_{j_2}^m(\cos\theta) = \delta_{j_1,j_2}. \quad (S8)$$

We use the orthogonal relations of Legendre polynomials and exponential functions in the integral transformation [4]. Firstly, consider the motion along rotational polar coordinate $\theta$. The product of two associated Legendre polynomials occur in Eq. S6 can be expanded by single associated Legendre polynomials

$$\widetilde{P}_{J_1}^{m_1}(\cos\theta)\widetilde{P}_{J_2}^{m_2}(\cos\theta) = \sum_{L=|J_1-J_2|}^{J_1+J_2} C_{J_1 m_1 J_2 m_2}^{L,m_1+m_2} \widetilde{P}_L^{m_1+m_2}(\cos\theta), \quad (S9)$$

$$C_{J_1 m_1 J_2 m_2}^{L,m_1+m_2} = \sqrt{\frac{(2J_1+1)(2J_2+1)}{4\pi(2L+1)}} \langle J_1 m_1 J_2 m_2 | L(m_1+m_2)\rangle \langle J_1 0 J_2 0 | L 0\rangle. \quad (S10)$$

Thus, integrate over $\theta$,

$$\begin{aligned}
I_{m_1 m_2}(\alpha, t) &= \int_0^\pi \sin\theta d\theta \widetilde{P}_\alpha^{m_1+m_2}(\cos\theta) \text{Pr}_{m_1,m_2}(\theta, t) \quad (S11) \\
&= \sum_{J_1=|m_1|}^\infty \sum_{J_2=|m_2|}^\infty \sum_{L=|\Delta J|}^{J} C_{J_1 m_1 J_2 m_2}^{L,m_1+m_2} \langle J_1 m_1 |\hat{\rho}| J_2 m_2\rangle e^{-i\Delta\omega t} \\
&\quad \times \int_0^\pi \sin\theta d\theta \widetilde{P}_\alpha^{m_1+m_2}(\cos\theta)\widetilde{P}_L^{m_1+m_2}(\cos\theta) \\
&= \sum_{J_1=|m_1|}^\infty \sum_{J_2=|m_2|}^\infty C_{J_1 m_1 J_2 m_2}^{\alpha,m_1+m_2} \langle J_1 m_1 |\hat{\rho}| J_2 m_2\rangle e^{-i\Delta\omega t}.
\end{aligned}$$

Let $T = 4\pi\mathcal{I}$, which is related to the rotational period, and integrate over $t$,

$$\begin{aligned}
I_{m_1 m_2}(\alpha, \beta) &= \frac{1}{T}\int_0^T I_{m_1 m_2}(\alpha, t) e^{i\beta(\alpha+1)t/2\mathcal{I}} dt \quad (S12) \\
&= \sum_{J_1=|m_1|}^\infty \sum_{J_2=|m_2|}^\infty C_{J_1 m_1 J_2 m_2}^{\alpha,m_1+m_2} \langle J_1 m_1 |\hat{\rho}| J_2 m_2\rangle \delta_{\beta(\alpha+1)-\Delta J(J+1)}.
\end{aligned}$$

The range of $\alpha$ and $\beta$ is set to be $|\Delta J| \leq |\beta| \leq \alpha \leq J$, where $\beta$ and $\Delta J$ are of the same sign. If $\beta(\alpha+1)$ has unique integer factorization, the only term remaining in the sum satisfying

$$\beta(\alpha+1) = \Delta J(J+1) \quad (S13)$$

6

is $\beta = \Delta J$ and $\alpha = J$. The corresponding density matrix element can be derived as

$$\langle \frac{\alpha+\beta}{2} m_1 | \hat{\rho} | \frac{\alpha-\beta}{2} m_2 \rangle = \frac{I_{m_1 m_2}(\alpha,\beta)}{C^{\alpha,m_1+m_2}_{\frac{\alpha+\beta}{2}m_1 \frac{\alpha-\beta}{2}m_2}} . \tag{S14}$$

If the factorization of $\beta(\alpha+1)$ is not unique, we calculate all integrations $I_{m_1 m_2}(\alpha',\beta')$ where $\beta(\alpha+1) = \beta'(\alpha'+1)$. For example, when $\beta = 0$,

$$I_{m_1 m_2}(\alpha, 0) = \sum_{J=\max\{|m_1|,|m_2|\}}^{\infty} C^{\alpha,m_1+m_2}_{Jm_1 Jm_2} \langle Jm_1 | \hat{\rho} | Jm_2 \rangle \tag{S15}$$

all of the $\Delta J = 0$ terms remain. When changing the value of $\alpha$, all these $I_{m_1 m_2}$ and corresponding density matrix elements constitute a set of linear algebraic equations (where $\alpha = 2J$ can only be even numbers),

$$\begin{pmatrix} I_{m_1 m_2}(\alpha, 0) \\ I_{m_1 m_2}(\alpha+2, 0) \\ I_{m_1 m_2}(\alpha+4, 0) \\ \vdots \end{pmatrix} = \begin{pmatrix} C^{\alpha,m_1+m_2}_{\frac{\alpha}{2}m_1 \frac{\alpha}{2}m_2} & C^{\alpha,m_1+m_2}_{\frac{\alpha}{2}+1,m_1,\frac{\alpha}{2}+1,m_2} & C^{\alpha,m_1+m_2}_{\frac{\alpha}{2}+2,m_1,\frac{\alpha}{2}+2,m_2} & \cdots \\ 0 & C^{\alpha+2,m_1+m_2}_{\frac{\alpha}{2}+1,m_1,\frac{\alpha}{2}+1,m_2} & C^{\alpha+2,m_1+m_2}_{\frac{\alpha}{2}+2,m_1,\frac{\alpha}{2}+2,m_2} & \cdots \\ 0 & 0 & C^{\alpha+4,m_1+m_2}_{\frac{\alpha}{2}+2,m_1,\frac{\alpha}{2}+2,m_2} & \cdots \\ \vdots & \vdots & \vdots & \cdots \end{pmatrix} \tag{S16}$$

$$\times \begin{pmatrix} \langle \frac{\alpha}{2} m_1 | \hat{\rho} | \frac{\alpha}{2} m_2 \rangle \\ \langle \frac{\alpha}{2}+1, m_1 | \hat{\rho} | \frac{\alpha}{2}+1, m_2 \rangle \\ \langle \frac{\alpha}{2}+2, m_1 | \hat{\rho} | \frac{\alpha}{2}+2, m_2 \rangle \\ \vdots \end{pmatrix},$$

which has unique solution because all diagonal terms of the upper triangular matrix are nonzero.

### 3. LASER ALIGNMENT OF ROTATING MOLECULE

The effective Hamiltonian of rotating molecule-laser interaction is [5]

$$\hat{H}_{\text{eff}} = \hat{H}_0 + \hat{H}_{\text{int}}$$
$$\hat{H}_0 = B\mathbf{J}^2$$
$$\hat{H}_{\text{int}} = -\frac{1}{2}\epsilon^2(t)[(\alpha_\parallel - \alpha_\perp)\cos^2\theta + \alpha_\perp], \tag{S17}$$

where $\mathbf{J}$ is the rotational angular momentum, $\epsilon(t)$ is the electric field of the laser pulse, $B$ is the rotational constant, $\alpha_\parallel$ and $\alpha_\perp$ are the components of the static polarizability, parallel

and perpendicular to the molecular axes. The molecule is assumed to be in the vibrational and electronic ground state. An initial rotational eigenstate $|J_0 M_0\rangle$ evolves to a pendular state [5]

$$|J_0 m_0\rangle \to |\psi(t)\rangle^{(J_0 m_0)} = \sum_J d_J^{(J_0 m_0)} |J m_0\rangle e^{-i E_J t/\hbar}, \quad (S18)$$

where $J$ and $J_0$ are of the same parity. The coupling coefficients $d_J^{J_0 m_0}$ is induced by laser field, satisfying selection rules $\Delta m = 0$ and $\Delta J = 0, \pm 2$. $d_J^{J_0 m_0}$ is invariant after the laser pulse, and the evolution of rotational angular distribution originates from interference of each dynamical phase. The coherence of the created quantum state can be maintained for several revival periods, and the alignment is reconstructed at predetermined times and survives for a perfectly controllable period [5], the sufficiently long coherence time makes the time evolution measurement of quantum state tomography feasible.

The initial system in thermal equilibrium can be characterized by the following density operator

$$\hat{\rho}_{\text{ini}} = \sum_{J_0 m_0} \omega_{J_0} |J_0 m_0\rangle\langle J_0 m_0|, \quad (S19)$$

where $\omega_{J_0}$ is the Boltzmann statistical factor determined by the rotational temperature. The density operator of the laser-aligned system is

$$\begin{aligned}
\hat{\rho}(t) &= \sum_{J_0 m_0} \omega_{J_0} |\psi(t)^{(J_0 m_0)}\rangle\langle \psi(t)^{(J_0 m_0)}| \quad (S20)\\
&= \sum_{m_0} \left[ \sum_{J_0} \omega_{J_0} \left( \sum_{J_1} d_{J_1}^{(J_0 m_0)} |J_1 m_0\rangle \right) \left( \sum_{J_2} d_{J_2}^{*(J_0 m_0)} \langle J_2 m_0| \right) \right] e^{-i(E_{J_1} - E_{J_2})t/\hbar} \\
&= \sum_{J_1 J_2 m} \left( \sum_{J_0} \omega_{J_0} d_{J_1}^{(J_0 m)} d_{J_2}^{*(J_0 m)} \right) e^{-i(E_{J_1} - E_{J_2})t/\hbar} |J_1 m\rangle\langle J_2 m|.
\end{aligned}$$

And its density matrix elements are

$$\langle J_1 m_1 | \hat{\rho}(t) | J_2 m_2 \rangle = \delta_{m_1 m_2} \left( \sum_{J_0} \omega_{J_0} d_{J_1}^{(J_0 m_1)} d_{J_2}^{*(J_0 m_2)} \right) e^{-i(E_{J_1} - E_{J_2})t/\hbar}. \quad (S21)$$

So the partial trace of $m$ subspace with odd (or even) $J$ is invariant in the dynamics of laser alignment, since it is a general property of laser-molecule interaction,

$$\sum_{J_{\text{odd}}} \langle J m | \hat{\rho} | J m \rangle = \sum_{J_{\text{odd}}} \sum_{J_{0\,\text{odd}}} \omega_{J_0} |d_J^{(J_0 m)}(t)|^2 = \sum_{J_{0\,\text{odd}}} \omega_{J_0}, \quad (S22)$$

where we used the normalization property of coefficients $d_J^{J_0 M}(t)$ in Eq. S18.

Notice that density matrix of opposite magnetic quantum number $m$ and $-m$ is symmetric for $\hat{\rho}_{\text{ini}}$, which also remains symmetric for transition matrix element induced by laser interaction $\hat{H}_{\text{eff}}(t)$. From Eq. S17, taking into account selection rule $\Delta M = 0$,

$$\langle J_1 m | \hat{H}_{\text{eff}}(t) | J_2 m \rangle = \langle J_1, -m | \hat{H}_{\text{eff}}(t) | J_2, -m \rangle$$
$$= \delta_{J_1,J_2}\left[BJ_1(J_1+1) - \frac{1}{2}\epsilon^2(t)\alpha_\perp\right] - \frac{1}{2}\epsilon^2(t)(\alpha_\parallel - \alpha_\perp)\langle J_1 m | \cos^2\theta | J_2 m \rangle, \quad \text{(S23)}$$

where $\langle J_1 m | \cos^2\theta | J_2 m \rangle = \langle J_1, -m | \cos^2\theta | J_2, -m \rangle$ according to the properties of Clebesh-Gordan coefficients. The coefficients of pendular state $d_J^{(J_0 m_0)}$, which are totally determined by initial condition $\hat{\rho}_{\text{ini}}$ and the Schördinger equation,

$$i\dot{d}_J^{(J_0 m)} = \sum_{J'} \langle Jm | \hat{H}_{\text{eff}}(t) | J'm \rangle, \quad \text{(S24)}$$

are also symmetric $d_J^{(J_0 m)} = d_J^{(J_0,-m)}$. So are the density matrix elements

$$\langle J_1 m_1 | \hat{\rho} | J_2 m_2 \rangle = \sum_{J_0} \omega_{J_0} d_{J_1}^{(J_0,m_1)} d_{J_2}^{*(J_0,m_2)} = \langle J_1, -m_1 | \hat{\rho} | J_2, -m_2 \rangle. \quad \text{(S25)}$$

## 4. THE ALGORITHM FOR IMPOSING CONSTRAINTS OF ITERATIVE QUANTUM TOMOGRAPHY

In this section we show the detailed procedure for making an arbitrary density matrix and probability distribution to satisfy the physical constraints given in the main text. Most physical constraints are given in the summation form. For example, from Eq. S22,

$$\sum_{J_{\text{odd}}} \langle Jm | \hat{\rho} | Jm \rangle = \sum_{J_0 \text{ odd}} \omega_{J_0}. \quad \text{(S26)}$$

From the measured probability distribution

$$\widetilde{\Pr}_{m_1-m_2}(\theta, t) = \frac{1}{2\pi}\int_0^{2\pi} d\phi \Pr(\theta, \phi, t) e^{-i(m_1-m_2)\phi} \quad \text{(S27)}$$
$$= \frac{1}{2\pi}\sum_{J_1 m_1' J_2 m_2'} \langle J_1 m_1' | \hat{\rho} | J_2 m_2' \rangle \widetilde{P}_{J_1}^{m_1}(\cos\theta)\widetilde{P}_{J_2}^{m_2}(\cos\theta) e^{-i\Delta\omega t}\int_0^{2\pi} d\phi e^{im_1'\phi}e^{-im_2'\phi}e^{-i(m_1-m_2)\phi}$$
$$= \sum_{m_1' m_2'} \delta_{m_1'-m_2',m_1-m_2}\sum_{J_1 J_2} \langle J_1 m_1' | \hat{\rho} | J_2 m_2' \rangle \widetilde{P}_{J_1}^{m_1}(\cos\theta)\widetilde{P}_{J_2}^{m_2}(\cos\theta) e^{-i\Delta\omega t},$$

9

and the constraint can be expressed as

$$\sum_{m'_1-m'_2=m_1-m_2} \Pr_{m'_1,m'_2}(\theta,t) = \widetilde{\Pr}_{m_1-m_2}(\theta,t). \tag{S28}$$

They can be sataisfied by scaling with a common factor

$$\langle Jm|\hat{\rho}|Jm\rangle \to \alpha \langle Jm|\hat{\rho}|Jm\rangle, \quad \alpha = \frac{\sum_{J_0\,\text{odd}} \omega_{J_0}}{\sum_{J_\text{odd}} \langle Jm|\hat{\rho}|Jm\rangle}. \tag{S29}$$

$$\Pr_{m_1,m_2}(\theta,t) \to \beta(\theta,t)\Pr_{m_1,m_2}(\theta,t), \quad \beta = \frac{\widetilde{\Pr}_{m_1-m_2}(\theta,t)}{\sum_{m'_1-m'_2=m_1-m_2} \Pr_{m'_1,m'_2}(\theta,t)}. \tag{S30}$$

The constraints in probability space is given by Eq. S30, and illustrated with flow chart in Fig. S2. Further constraints in density matrix space include being Hermitian, positive semidefinite and having invariant partial traces (the procedure is presented with the flow chart in Fig. S3).

As a general rule to guarantee the completeness of constraint conditions, we can firstly analyse the physical system and find out the possible states, which could give same probability distribution for all time and are indistinguishable without further constraint, and construct the set of physical conditions that can distinguish the states from each other, e.g. selection rules, symmetry. The obtained physical conditions can be then used as constraints in the iterative QT procedure. In this manner, the completeness of the constraint conditions and the faithfulness of the converged density matrix solution can be achieved, i.e. the converged solution of the inversion problem is the true density matrix of the physical system.

## 5. NUMERICAL TRIAL WITH RANDOMLY CHOSEN DENSITY MATRIX AND INITIAL GUESS

We have verified the new quantum tomographic method by the rotational wavepacket of a laser-aligned molecule. We also illustrate the power of the new method by applying it to a randomly chosen density matrix rather than that in the laser-aligned case. The iterative QT algorithm also converges after about 20 iterations and density matrix is recovered with

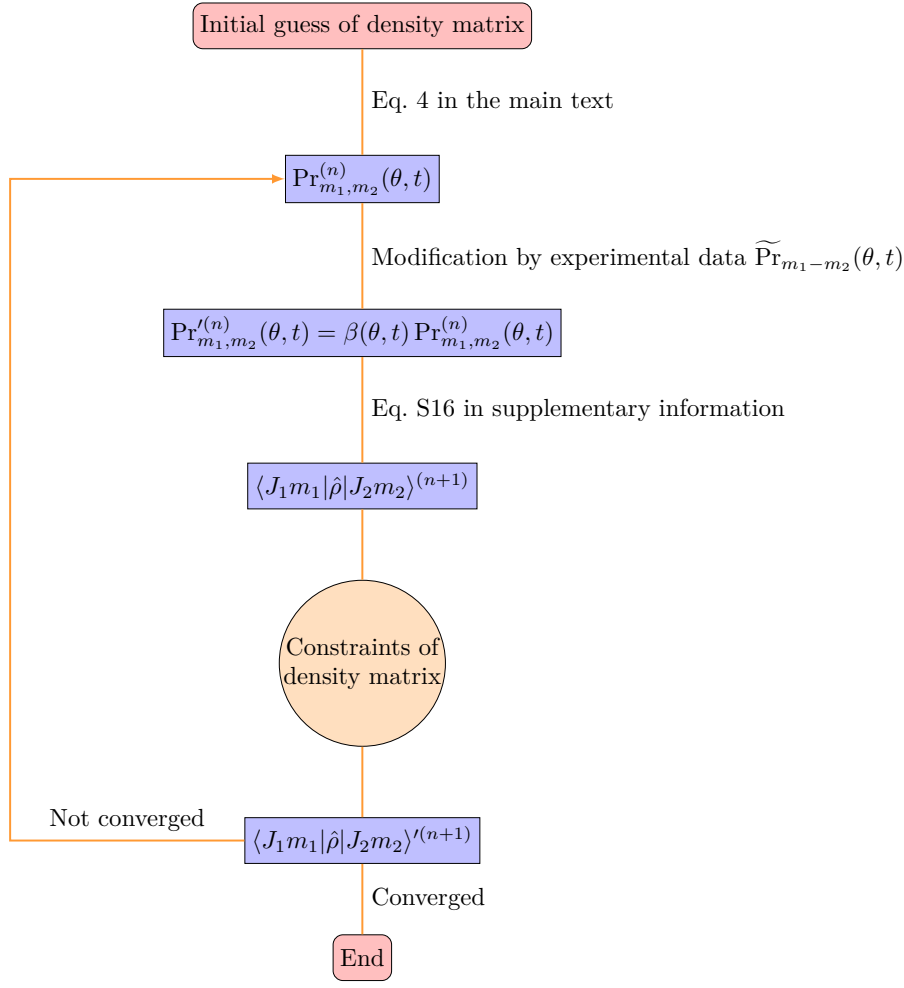

FIG. S2. Schematic flow chart for imposing constraints to the wavepacket probability distribution. The internal procedure for the "constraints of density matrix" is separately elaborated in Fig. S3. The superscript $n$ represents $n$-th iteration.

considerable accuracy. The density operator of the state to be recovered is set to be

$$\begin{aligned}
\hat{\rho} =\ & \frac{2}{21}|00\rangle\langle 00| + \frac{3}{14}|10\rangle\langle 10| + \frac{1}{42}|20\rangle\langle 20| \\
& + \left(\frac{1}{7}|00\rangle\langle 10| + \frac{1}{21}|00\rangle\langle 20| + \frac{1}{14}|10\rangle\langle 20| + \text{H.c.}\right) \\
& + \frac{2}{21}|11\rangle\langle 11| + \frac{3}{14}|21\rangle\langle 21| + \frac{1}{42}|31\rangle\langle 31| \\
& + \left(\frac{1}{7}|11\rangle\langle 21| + \frac{1}{21}|11\rangle\langle 31| + \frac{1}{14}|21\rangle\langle 31| + \text{H.c.}\right) \\
& + \frac{2}{21}|22\rangle\langle 22| + \frac{3}{14}|32\rangle\langle 32| + \frac{1}{42}|42\rangle\langle 42| \\
& + \left(\frac{1}{7}|22\rangle\langle 32| + \frac{1}{21}|22\rangle\langle 42| + \frac{1}{14}|32\rangle\langle 42| + \text{H.c.}\right).
\end{aligned} \qquad (S31)$$

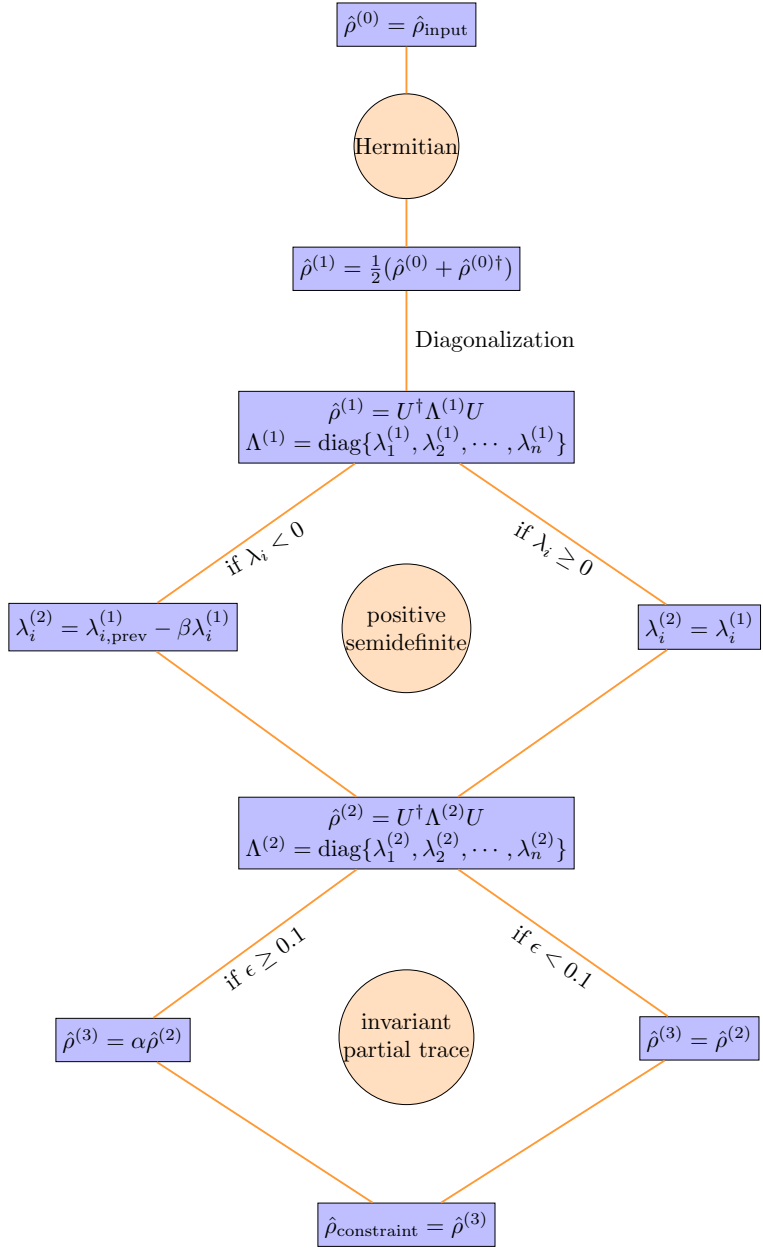

FIG. S3. Schematic flow chart for imposing constraints to the density matrix, where $\epsilon = \left| \frac{\sum_{J_{\text{odd}}} - \sum_{J_{0\,\text{odd}}} \omega_{J_0}}{\sum_{J_{0\,\text{odd}}} \omega_{J_0}} \right|$. $\alpha$ is defined in Eq. S29. We use hybrid input-output (HIO) algorithm for the positivity constraint with $\beta = 0.9$ [3], where the subscript "prev" stands for the use of values in the previous iteration.

We impose the error functions of density matrix and probability distribution to measure

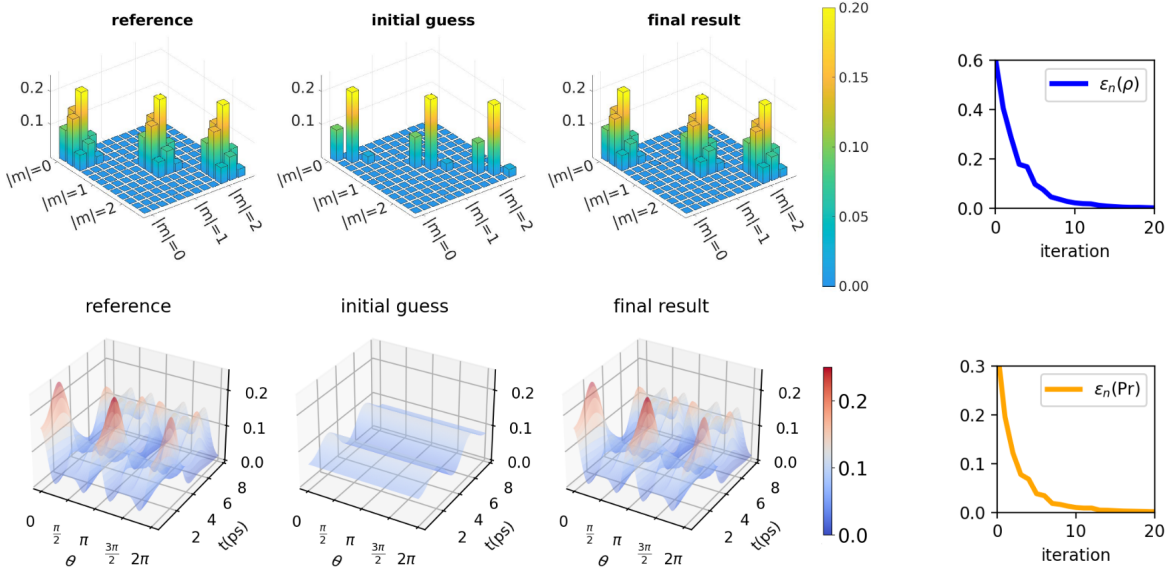

FIG. S4. Quantum tomography result of numerical trial with initial guess of thermal equilibrium state. The modulus of density matrix elements are shown in the upper panel, where $J = |m|, |m|+1, \cdots, J_{\max}$ within each $m$-block. The lower panel shows angular probability distribution, the recovered modulus and phases of density matrix elements faithfully reproduce the reference $\Pr(\theta, t)$. Error functions of density matrix and probability distribution are shown in the rightmost column.

the accuracy of iteration results, which are defined by

$$\epsilon_n(\hat{\rho}) = \frac{\sum_{J_1 m_1 J_2 m_2} |\langle J_1 m_1|\hat{\rho}|J_2 m_2\rangle_n - \langle J_1 m_1|\hat{\rho}|J_2 m_2\rangle_0|}{\sum_{J_1 m_1 J_2 m_2} |\langle J_1 m_1|\hat{\rho}|J_2 m_2\rangle_0|} \tag{S32}$$

$$\epsilon_n(\Pr) = \frac{\sum_{i,j,k} |\Pr_n(\theta_i, \phi_j, t_k) - \Pr_0(\theta_i, \phi_j, t_k)|}{\sum_{i,j,k} |\Pr_0(\theta_i, \phi_j, t_k)|} \tag{S33}$$

where the subscript $n$ represents the result of $n$-th iteration, and 0 represents the correct result.

In Fig. S4 we show the result of identical algorithm given in Fig. S2 and Fig. S3, only with smaller order $J_{\max}$ of density matrix to be recovered. The thermal equilibrium initial state is given by the rotational temperature, which can be measured in the experiment [6, 7]

$$\hat{\rho}_{\mathrm{ini}} = \sum_{J_0 m_0} \omega_{J_0} |J_0 m_0\rangle\langle J_0 m_0|, \tag{S34}$$

where $\omega_{J_0}$ is the Boltzmann statistical weight. The iteration converged to the expected result with error $\epsilon_{20}(\hat{\rho}) = 3.5 \times 10^{-3}$ and $\epsilon_{20}(\Pr) = 1.7 \times 10^{-3}$.

13

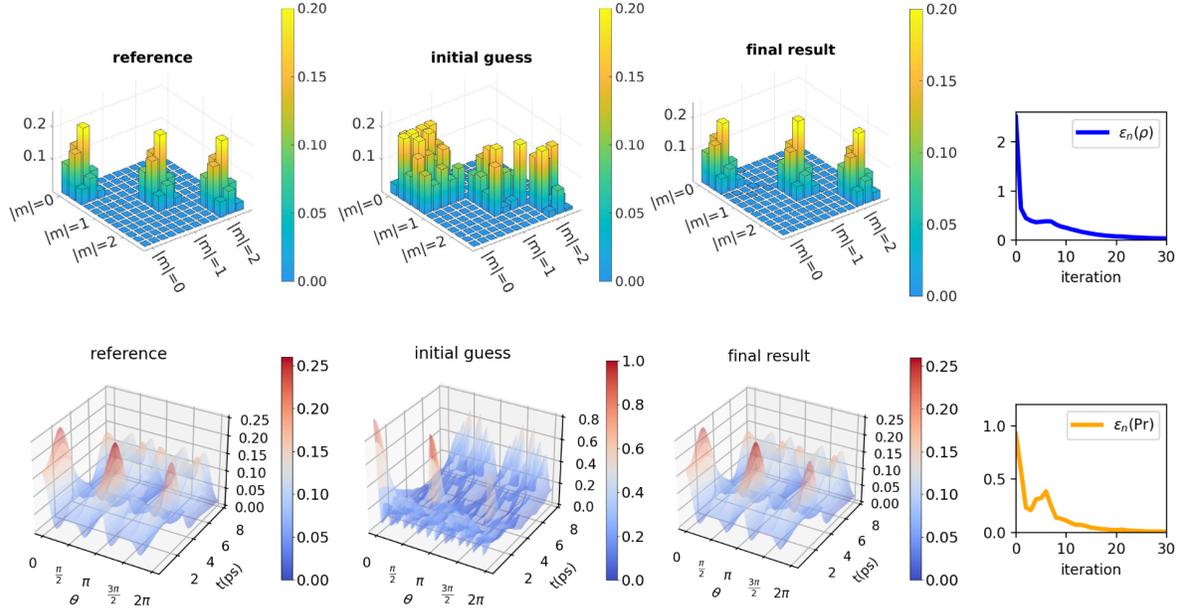

FIG. S5. Quantum tomography result of numerical trial with random initial guess of density matrix. Only the measured probability distribution and general properties of density matrix (being Hermitian, positive semidefinite and with unity trace) are imposed as constraints during the iteration algorithm. The density matrix to be recovered and its probability distribution are identical to that in Fig. S4. The modulus of density matrix elements are shown in the upper panel, where $J = |m|, |m|+1, \cdots, J_{\max}$ within each $m$-block. The lower panel shows angular probability distribution, the recovered modulus and phases of density matrix elements faithfully reproduce the reference $\Pr(\theta, t)$. Error functions of density matrix and probability distribution are shown in the rightmost column.

Especially, we show with the proof-of-principle example that this iterative QT algorithm is insensitive with the initial guess of density matrix. The rotational temperature which provides much information such as initial guess and partial trace, is actually not indispensable to the QT method. Assume we are dealing with a pure QT problem without any additional knowledge to the density matrix to be recovered. As is shown in Fig. S5, a random initial guess will also lead to a converged result after about 30 iterations with error $\epsilon_{30}(\hat{\rho}) = 3.9 \times 10^{-2}$ and $\epsilon_{30}(\Pr) = 9.0 \times 10^{-3}$.

14

## 6. VIBRATIONAL QUANTUM TOMOGRAPHY

Vibrational quantum tomography recovers the density matrix of $N$ vibrational modes from the probability distribution evolution $\Pr(x_1, x_2, \cdots, x_N, t)$

$$\Pr(x_1, x_2, \cdots, x_N, t) = \sum_{\{m_i\}_{i=1}^N} \sum_{\{n_i\}_{i=1}^N} \langle n_1 n_2 \cdots n_N | \hat{\rho} | m_1 m_2 \cdots m_N \rangle \tag{S35}$$

$$\times \prod_{i=1}^N \phi_{n_i}(x_i) \phi_{m_i}^*(x_i) e^{i(m_i - n_i)\omega_i t}.$$

The dimension problem arises naturally here. In conventional QT method that is based on integral transform, the orthogonal properties cancel out one summation by integrating over one parameter. For example,

$$\frac{1}{T} \int_0^T dt e^{i(m-n)r\omega_0 t} e^{-ik\omega_0 t} = \delta_{(m-n)r,k}, \tag{S36}$$

where $T = \frac{2\pi}{\omega_0}$. $f_{mn}(x)$ is the sampling function [8] defined by

$$f_{mn}(x) = \frac{\partial}{\partial x}[\phi_m(x)\varphi_n(x)], \tag{S37}$$

where $\phi_m(x)$ and $\varphi_n(x)$ are respectively regular and irregular wavefunctions of harmonic oscillator. The bi-orthogonal properties of sampling function is

$$\int_{-\infty}^{+\infty} dx f_{mn}(x) \phi_{m'}^*(x) \phi_{n'}(x) = \delta_{mm'} \delta_{nn'}, \tag{S38}$$

under frequency constraints $m - n = m' - n'$.

Here the probability is $(N+1)$-dimensional and density matrix is $2N$-dimensional, which is inadmissible for analytical solutions when $N > 1$.

Our theory, based on the following two procedures, fully utilizes the above orthogonal properties and imposes constraints for lack of dimension. First, we set up the transformation between probability and density matrix in a subspace

$$\Pr\nolimits_{\Delta_1, \Delta_2, \cdots, \Delta_N}(x_1, x_2, \cdots, x_N) = \sum_{\{m_i\}_{i=1}^N} \sum_{\{n_i\}_{i=1}^N} \langle n_1 n_2 \cdots n_N | \hat{\rho} | m_1 m_2 \cdots m_N \rangle \tag{S39}$$

$$\times \prod_{i=1}^N \phi_{n_i}(x_i) \phi_{m_i}^*(x_i) \delta_{m_i - n_i, \Delta_i}$$

$$\langle n_1 n_2 \cdots n_N | \hat{\rho} | m_1 m_2 \cdots m_N \rangle = \int d^N \mathbf{x} \Pr\nolimits_{\Delta_1, \Delta_2, \cdots, \Delta_N}(x_1, x_2, \cdots, x_N) \prod_{i=1}^N f_{m_i n_i}(x_i). \tag{S40}$$

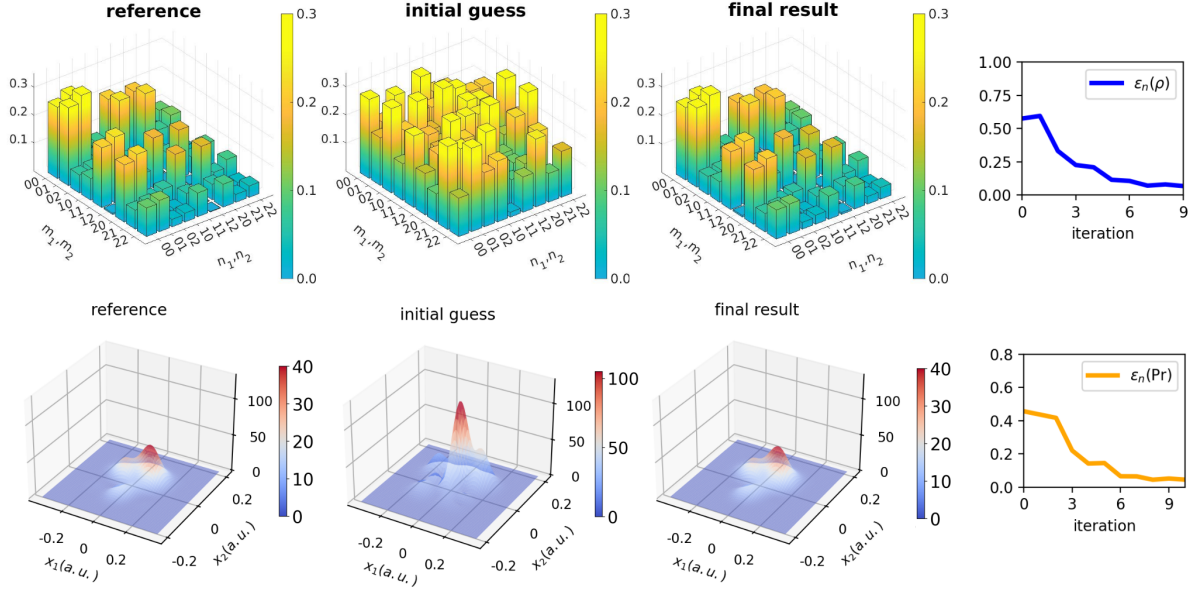

FIG. S6. Quantum tomography of two dimensional vibrational wavepacket to the second order, with reduced mass 12amu, frequency $\omega_0 = 1209.8 \text{cm}^{-1}$ (0.15eV) and frequency ratio of two vibrational modes $r_1/r_2 = 1/3$. The modulus of density matrix elements and probability distribution for a given time $t = 1.8\text{fs}$ are shown in the upper panel and lower panel, the recovered modulus and phases of density matrix elements faithfully reproduce the reference $\Pr(x_1, x_2, t)$. The algorithm converged for about 10 iterations as illustrated in the rightmost column, where $\epsilon_{10}(\hat{\rho}) = 4.1 \times 10^{-2}$ and $\epsilon_{10}(\Pr) = 3.1 \times 10^{-2}$.

Second, starting from an initial guess, effective physical constraints can be imposed by iterative projection method to get the converged result. For example, the a priori knowledge of density matrix of being Hermitian, positive semidefinite and normalized. An example of 2D vibrational quantum tomography is shown in Fig. S6. The initial guess is given randomly, and only the probability distribution and general properties of density matrix are imposed as constraints during the iteration algorithm.

Similar to rotational QT, the dimension problem can be reflected by the fact that for

$$\Pr_k(x_1, x_2, \cdots, x_N) = \sum_{\{\Delta_i\}_{i=1}^N} \Pr_{\Delta_1, \Delta_2, \cdots, \Delta_N}(x_1, x_2, \cdots, x_N) \delta_{\sum_{i=1}^N \Delta_i r_i, k}, \tag{S41}$$

unless only one single combination of $\{\Delta_i\}$ satisfies $\sum_{i=1}^N \Delta_i r_i = k$, there is no direct way to obtain $\Pr_{\Delta_1, \Delta_2, \cdots, \Delta_N}(x_1, x_2, \cdots, x_N)$ from the measured wavepacket density distribution, only their sum can be available through Fourier transform of the measured probability

16

distribution evolution

$$\Pr{}_k(x_1, x_2, \cdots, x_N) = \frac{1}{T} \int_0^T dt\, e^{-ik\omega_0 t} \Pr(x_1, x_2, \cdots, x_N, t), \tag{S42}$$

where we assume $\omega_i = r_i \omega_0$ ($r_i$ are integers and $T = 2\pi/\omega_0$, $r_i$'s are the set of smallest integers to represent the measured frequencies).

In the new iterative QT method for $N$-dimensional vibrational system, we do not need infinitely long time of measurement anymore, which used to be indispensable to fill the whole space of $N$-dimensional phases [9] while physically infeasible. Besides, in the new iterative QT method, the ratio of frequencies does not have to be irrational, which is important because in reality $N$-dimensional vibrational systems with commensurable frequencies are ubiquitous.

The pattern function can be approximated around $x = 0$ as [10]

$$f_{nn} \sim -\frac{2}{\pi} \sin[-\pi(n + 1/2) + 2\sqrt{2n + 1}x]. \tag{S43}$$

In order to resolve a period of the oscillation of the pattern function that arises in the convolution (Eq. S40), the required spatial resolution for reconstructing vibrational density matrix up to $N$th order has to be better than $\delta x \leq \pi/2\sqrt{2N+1}$. The maximal order of the desired density matrix also sets demand on the temporal resolution. Suppose $d$ time intervals are measured for a half period $T/2 = \pi/\omega_0$. From Eq. S42, we have a phase resolution of $k\pi/d$ for the Fourier transformation of probability distribution function. The aliasing phenomena defines the maximal order of density matrix we can access to be $N = d/k - 1$, thus the required temporal resolution is

$$\delta t \leq \frac{T}{2(N+1)k} \leq \frac{T}{2(N+1)\sum_i r_i}. \tag{S44}$$

The quantum tomography procedure presented above can be easily generalized to systems when coupling among different vibrational modes exist. In general case, the Hamiltonian [11]

$$\hat{H} = \sum_{i=1}^{N} \hat{h}_i + V(x_1, x_2, \cdots, x_N), \tag{S45}$$

where $\hat{h}_i$ is the separable part for $i$-th vibrational mode with eigenstate $\phi_{n_i}(x_i)$, and $V(x_1, x_2, \cdots, x_N)$ is coupling potential among $N$ vibrational modes. The eigenstate is

17

a linear combination of product 1D wavefunctions assigned with quantum numbers $I = \{I_1, I_2, \cdots, I_N\}$ with energy eigenvalue $E_I$

$$\Psi_I(x_1, x_2, \cdots, x_N) = \sum_{i_1, i_2, \cdots, i_N} C_I^{i_1, i_2, \cdots, i_N} \prod_{\alpha=1}^{N} \phi_{i_\alpha}(x_\alpha). \tag{S46}$$

The iterative projection algorithm for quantum tomography should be set up based on the transformation between probability and density matrix in a subspace

$$\Pr_{\Delta_1, \Delta_2, \cdots, \Delta_N}(x_1, x_2, \cdots, x_N) = \sum_{I,J} \langle I|\hat{\rho}|J\rangle \sum_{i_1, i_2, \cdots, i_N} \sum_{j_1, j_2, \cdots, j_N} C_I^{i_1, i_2, \cdots, i_N} C_J^{j_1, j_2, \cdots, j_N *} \tag{S47}$$
$$\times \prod_{\alpha=1}^{N} \phi_{i_\alpha}(x_\alpha) \phi_{j_\alpha}^*(x_\alpha) \delta_{i_\alpha - j_\alpha, \Delta_\alpha}$$

$$\int d^N \mathbf{x} \Pr_{\Delta_1, \Delta_2, \cdots, \Delta_N}(x_1, x_2, \cdots, x_N) \prod_{\alpha=1}^{N} f_{i_\alpha j_\alpha}(x_\alpha) = \sum_{I,J} \langle I|\hat{\rho}|J\rangle C_I^{i_1, i_2, \cdots, i_N} C_J^{j_1, j_2, \cdots, j_N *}. \tag{S48}$$

where the frequency constraint of sampling function requires $i_\alpha - j_\alpha = \Delta_\alpha$ ($\alpha = 1, 2, \cdots, N$). The density matrix element can be solved from the linear equation of S48. If there are $n$ basis eigenstate for $i$-th uncoupled vibrational mode $\phi_{n_i}(x_i)$, the coupled density matrix can be recovered to the order of $(2n)^{N/2}$. Similarly, the procedure starts from an initial guess and imposes constraints to both density matrix space and probability space. Besides basic properties of density matrix and probability distribution, the subspace probability should also satisfy

$$\Pr_{\omega_{IJ}}(x_1, x_2, \cdots, x_N) = \frac{1}{T} \int_0^T dt \Pr(x_1, x_2, \cdots, x_N, t) e^{-i\omega_{IJ} t} \tag{S49}$$
$$= \sum_{E_I - E_J = \omega_{IJ}} \langle I|\hat{\rho}|J\rangle \varphi_{i_1, i_2, \cdots, i_N}(x_1, x_2, \cdots, x_N) \varphi_{j_1, j_2, \cdots, j_N}^*(x_1, x_2, \cdots, x_N)$$
$$= \sum_{\Delta_1, \Delta_2, \cdots, \Delta_N} \Pr_{\Delta_1, \Delta_2, \cdots, \Delta_N}(x_1, x_2, \cdots, x_N) \delta_{E_I - E_J, \omega_{IJ}}.$$

where $E_I$ and $E_J$ are energy eigenvalues of the coupled Hamiltonian, $T$ is the common period for all vibrational frequency intervals.

To enhance the convergence of iterative QT procedure for vibrational states, physical constraints can be imposed on the diagonal matrix elements of the density matrix, which is experimentally accessible, e.g. through photoelectron spectra and absorption spectra, which can directly provide constraints on diagonal density matrix elements of basis states with eigenenergy $E$ [12].

As a final remark, for vibrational QT, it is sometimes neccessary to use the velocities of nuclei as constraining physical conditions, in the case that the basis states of density matrix is energetically degenerate. For example, given the ratio of two vibrational frequencies $r_1/r_2 = 1/2$, consider a mixed state consisting of $|20\rangle$ and $|10\rangle$ (the pure state is a special case of it), their density matrix is

$$\rho = \begin{pmatrix} \langle 20|\hat{\rho}|20\rangle & \langle 20|\hat{\rho}|01\rangle \\ \langle 01|\hat{\rho}|20\rangle & \langle 01|\hat{\rho}|01\rangle \end{pmatrix} = \begin{pmatrix} \rho_{11} & \rho_{12} \\ \rho_{21} & \rho_{22} \end{pmatrix}. \tag{S50}$$

The probability distribution

$$\begin{aligned} \Pr(x_1, x_2, t) &= \rho_{11}\phi_2^2(x_1)\phi_0^2(x_2) + \rho_{22}\phi_0^2(x_1)\phi_1^2(x_2) \\ &+ (\rho_{12} + \rho_{21})\phi_2(x_1)\phi_0(x_2)\phi_0(x_1)\phi_1(x_2) \end{aligned} \tag{S51}$$

could not reflect the imaginary part of the off-diagonal density matrix elements because the degeneracy of the two basis states smears out the temporal evolution of the probability distribution. If $|20\rangle$ and $|01\rangle$ belong to the same symmetry representation, their coupling will lead to Fermi resonance and the degeneracy can be lifted. In the case that $|20\rangle$ and $|01\rangle$ are exactly degenerate, additional constraints must be imposed. Because with the ultrafast diffraction method, the velocity of nuclei and thus thus their momenta can be extracted experimentally, we can naturally construct physical constraints through products of momenta, such as $p_{x_1}^2 p_{x_2}$, since

$$\mathbf{A} = (\hat{p}_{x_1}^2 \hat{p}_{x_2}) = \begin{pmatrix} a_{11} & a_{12} \\ a_{21} & a_{22} \end{pmatrix} \tag{S52}$$

has nonzero imaginary part of non-diagonal matrix elements. For example,

$$\begin{aligned} a_{12} &= \int dx_1 \phi_2(x_1) \left(-\frac{\partial^2}{\partial x_1^2}\right) \phi_0(x_1) \int dx_2 \phi_0(x_2) \left(-i\frac{\partial}{\partial x_2}\right) \phi_1(x_2) \\ &= \int_{-\infty}^{\infty} dx_1 \frac{1}{\pi^{1/4}} \sqrt{\frac{\alpha_1}{2}} (2\alpha_1^2 x_1^2 - 1) e^{-\frac{1}{2}\alpha_1 x_1^2} \frac{\partial^2}{\partial x_1^2} \left(\frac{\sqrt{\alpha_1}}{\pi^{1/4}} e^{-\frac{1}{2}\alpha_1 x_1^2}\right) \\ &\times \int_{-\infty}^{\infty} dx_2 \frac{\sqrt{\alpha_2}}{\pi^{1/4}} e^{-\frac{1}{2}\alpha_2 x_2^2} \frac{\partial}{\partial x_2} \left(\frac{\sqrt{2\alpha_2}}{\pi^{1/4}} e^{-\frac{1}{2}\alpha_2 x_2^2}\right) = -i\frac{\alpha_1^2 \alpha_2}{2} \\ a_{21} &= a_{12}^* = i\frac{\alpha_1^2 \alpha_2}{2} \end{aligned} \tag{S53}$$

19

The observable

$$\begin{aligned}\langle \hat{A}\rangle &= m_1^2 v_1^2 m_2 v_2 = \text{Tr}(\hat{\rho}\hat{A}) \quad &\text{(S54)}\\ &= \rho_{11}a_{11} + \rho_{12}a_{21} + \rho_{21}a_{12} + \rho_{22}a_{22}\\ &= \rho_{11}a_{11} + \rho_{22}a_{22} + 2\text{Re}[\rho_{12}a_{21}]\\ &= \rho_{11}a_{11} + \rho_{22}a_{22} - \alpha_1^2\alpha_2 \text{Im}[\rho_{12}]\end{aligned}$$

contains information of imaginary part of non-diagonal density matrix elements $\text{Im}[\rho_{12}] = -\text{Im}[\rho_{21}]$, with which we can effectively determine the imaginary part of the off-diagonal density matrix elements between exactly degenerate basis states, by using the products of velocities as physical constraints in the iterative QT procedure.

Throughout the paper, we focus on recovering the density matrix, which is interconnected with the Wigner function $W(q,p)$ via the overlapping formula,

$$\begin{aligned}\rho_{mn} &= \text{Tr}[\hat{\rho}|n\rangle\langle m|]\\ &= \frac{1}{2\pi}\int_{-\infty}^{\infty} dq \int_{-\infty}^{\infty} dp\, W(q,p) W_{|n\rangle\langle m|}(q,p),\end{aligned} \quad \text{(S55)}$$

where $W_{\hat{\mathcal{O}}}(q,p) = (1/2\pi)\int dx \exp(-ipx)\langle q - \frac{x}{2}|\hat{\mathcal{O}}|q+\frac{x}{2}\rangle$. Especially, the Wigner function can be expressed in terms of the density operator $\hat{\rho}$ as $W(q,p) = W_{\hat{\rho}}(q,p)$.

---



\end{document}